\documentclass[twocolumn]{autart} 
\usepackage{generic}
\usepackage{cite}
\usepackage{amsmath,amssymb,amsfonts}
\usepackage{subcaption}
\usepackage{algorithmic}

\usepackage{algorithm}
\usepackage[normalem]{ulem}
\usepackage{graphicx}
\usepackage{wrapfig}
\usepackage{textcomp}
\usepackage{cite}
\usepackage{textcomp}
\usepackage{setspace}
\usepackage{multirow}
\usepackage{xcolor}
\usepackage{bbding}
\usepackage{booktabs}
\usepackage{threeparttable}
\usepackage{bm}
\def\BibTeX{{\rm B\kern-.05em{\sc i\kern-.025em b}\kern-.08em
 T\kern-.1667em\lower.7ex\hbox{E}\kern-.125emX}}

\usepackage{threeparttable}
\allowdisplaybreaks
\begin{document}
\begin{frontmatter}
\title{Distributed { Safe} Control Design and {Probabilistic} Safety Verification for Multi-Agent Systems\thanksref{footnoteinfo}} % Title, preferably not more   % than 10 words.
\thanks[footnoteinfo]{For the purpose of Open Access, the authors have applied a CC BY public copyright licence to any Author Accepted Manuscript (AAM) version
arising from this submission.
Part of the results of this manuscript has been presented at the IEEE Conference on Decision and Control 2023 \cite{cdcsubmit}. Here we significantly extend the conference version by additionally evaluating a lower bound on the probability of safety in Section \ref{sec:safetyverification}, proposing a distributed safe controller design algorithm, and a truncated algorithm with rigorous safety analysis in Section \ref{sec:dscl}.\\
Antonis Papachristodoulou was supported in part by the EPSRC Programme Grant EEBio (EP/Y014073/1). Han Wang was supported by the EPSRC IAA Technology Fund. Kostas Margellos would like to acknowledge partial support by MathWorks.}

\author{Han Wang}\ead{han.wang@eng.ox.ac.uk},  % Add the 
\author{Antonis Papachristodoulou}\ead{antonis@eng.ox.ac.uk},        % e-mail address 
\author{Kostas Margellos}\ead{kostas.margellos@eng.ox.ac.uk} % (ead) as shown

\address{OX1 3PJ, Department of Engineering Science, University of Oxford, Oxford, United Kingdom.} % Please supply       

\maketitle

\begin{abstract}
We propose distributed iterative algorithms for safe control design and safety verification for networked multi-agent systems. These algorithms rely on distributing a control barrier function (CBF) related quadratic programming (QP) problem {assuming the existence of CBFs.} The proposed distributed algorithm addresses infeasibility issues of existing schemes {via a cooperation mechanism between agents}. The resulting control input is guaranteed to be optimal, and {satisfies CBF constraints of all agents}. Furthermore, a truncated algorithm is proposed to facilitate computational implementation. {The performance of the truncated algorithm is evaluated using a distributed safety verification algorithm. The algorithm quantifies safety for multi-agent systems probabilistically by means of CBFs.} Both upper and lower bounds on the probability of safety are obtained using the so called scenario approach. Both the scenario sampling and safety verification procedures are fully distributed. The efficacy of our algorithms is demonstrated by an example on multi-robot collision avoidance.
\end{abstract}

\begin{keyword}
Distributed Control, Safe Control, Multi-Agent Systems, Scenario Approach, Nonlinear Systems
\end{keyword}
\end{frontmatter}

\section{Introduction}
\label{sec:intro}
Safety of a dynamical system requires the system state to remain in a safe set for all time. This property is important in many applications such as collision avoidance \cite{ding2022configuration,wang2019moving}, vehicle platooning \cite{axelsson2016safety,alam2014guaranteeing}, vehicle merging control \cite{xiao2021decentralized}, etc. For a single agent system, safety is usually captured by introducing constraints on the state of the agent and the environment. For a multi-agent system, the meaning of safety extends to capture the interactions among agents. In this case, safety is encoded by coupling constraints over the states of a group of agents. For a networked multi-agent system, where agents cooperate to satisfy safety constraints, we consider designing distributed algorithms to ensure safety for all agents. 

Another problem of interest is to validate the proposed control law. For a single agent system, an agent can evaluate the system behaviour to characterize its risk of being unsafe under the employed control input. Similarly, for a multi-agent safety verification problem, cooperation among agents is necessary since safety involves multiple agents. In summary, this paper focuses on designing a distributed protocol for safe control input design and developing a distributed safety verification algorithm.

\subsection{Related Work}
Safety in control systems is often certified by control barrier functions (CBF), which is a type of control Lyapunov-like functions \cite{ames2016control,sontag1989universal,primbs1999nonlinear}. By enforcing the inner product of the CBF derivative and vector field of the controlled system to be bounded, safety is rigorously guaranteed at any time. CBF is shown to be powerful and scalable in control input design for control-affine systems, as this condition can be encoded as a linear constraint in a quadratic programming (QP) problem \cite{ames2016control}. By solving online QP problems at every state, the system is guaranteed to be safe \cite{hsu2015control,ames2014control}. 
% Higher-order derivative based methods for high relative degree systems are proposed in \cite{xiao2019control,nguyen2016exponential,tan2021high}. In \cite{xiao2021adaptive}, adaptive coefficients are introduced to improve the feasibility of the CBF-QP. For the case where multiple CBFs exist, an optimal decay based method is proposed to tune the CBF constraints \cite{zeng2021safety}. CBFs for discrete time systems are proposed in \cite{agrawal2017discrete}. For the case where model uncertainty and system noise are added, robust CBF with worst case analysis \cite{nguyen2021robust, jankovic2018robust} can be considered. 
Most of the existing results in this direction involve a centralized approach; however, multi-agent considerations call for distributed solution regimes.

{CBF-based distributed algorithms have been proposed in \cite{chen2020guaranteed,wang2017safety,borrmann2015control}.
In these papers, the CBF constraints are decomposed and allocated to neighbouring agents to facilitate a distributed implementation. Under the assumption that each local optimization problem is feasible, the overall CBF constraints are satisfied. However, this assumption is usually much stronger than that of feasibility of the nominal centralized problem. Moreover, optimality of the nominal centralized problem by the distributed controller is not guaranteed.} An improved constraint sharing mechanism is developed in \cite{xu2018constrained}, where the CBF constraints are dynamically tuned for feasibility, but for single-agent systems. Optimality is further considered in \cite{tan2021distributed} {, but for multi-agent systems with only one CBF constraint}. A dynamical constraint allocation scheme among agents based on a consensus protocol is proposed. {In our work, we deal with the problem of guaranteeing feasibility of local problems across iterations while preserving optimality, under multiple CBF safety constraints.} In essence, the distributed CBF-based safe control design problem can be seen under the lens of distributed optimization. 

Distributed optimization for a multi-agent system aims to design a distributed protocol that involves solving an optimization problem locally for every agent. Algorithms can be divided into two types, dual decomposition \cite{falsone2017dual,falsone2020tracking,shi2014linear,duchi2011dual} and primal decomposition-based \cite{margellos2017distributed,camisa2021distributed,notarnicola2019constraint,li2020distributed,nedic2010constrained}. Dual decomposition methods consider the dual problem, where each agent maintains a local copy of the dual variables. Constraint satisfaction is achieved by consensus over the dual variables. Primal decomposition methods directly decompose the primal problem into local problems. By local projection \cite{margellos2017distributed,li2020distributed,nedic2010constrained} or updating auxiliary variables \cite{camisa2021distributed,notarnicola2019constraint}, algorithms converge to centralized optimum under convexity assumptions. Such methods guarantee near feasibility as far as the constraints of the primal problem are concerned. As our problem has similar structure as the one considered in \cite{camisa2021distributed,notarnicola2019constraint}, primal decomposition structure is applied to develop our algorithm. 

To reduce the communication and computation burden, a truncation mechanism is proposed to allow us to terminate the algorithm before reaching convergence. To give a probabilistic guarantee for safety over the state space, we leverage scenario approach \cite{campi2008exact,campi2018wait,calafiore2005uncertain,calafiore2006scenario,garatti2019risk}, which samples a number of independent states from the state space and enforces the constraint only at these realizations.

% Another problem of interest in this work is safety verification. For a dynamical system, safety requires the trajectory to be within a safe set. Given the vector field, a target set and an unsafe set, solving a reach-avoid game \cite{margellos2011hamilton,lygeros2004reachability} yields a set from which all trajectories start can reach the target set without entering the unsafe set. In this sense, safety verification lies in the scope of reachability analysis. The main challenge here is how to solve the underlying Hamilton Jacobi partial differential equation. To bypass this difficulty, the barrier certificates method was proposed in a convex programming framework~\cite{prajna2004safety,prajna2007framework}. A barrier certificate identifies an invariant set inside the safe set. System trajectories cannot escape from the underlying invariant set, and this directly leads to safety. Numerical methods for verifying safety using barrier certificates with convex programs entails sum-of-squares (SOS) programs~\cite{prajna2002introducing,prajna2004sostools}, which are equivalent to semi-definite programs. In real applications, the system model and control input are usually not precisely known, or are even unknown. In this setup, another type of verification method \cite{akella2022barrier} using sampled data was proposed recently using scenario approach, but for systems with a single agent. {We extend this setting to the multi-agent problem which in turn imposes additional challenges.}

{
\subsection{Contributions}
Our contributions can be summarized as follows:
\begin{enumerate} 
\item We provide a distributed algorithm for designing safe controllers for multi-agent systems. Under the assumption of the existence of {feasible} CBFs, a centralized safe control design problem is formulated. Our distributed algorithm parallelizes computation by decomposing the centralized problem into local problems, while guaranteeing feasibility of every local problem across iterations. The optimal solution returned by our algorithm is guaranteed to be the same as that of the nominal centralized problem, therefore satisfying all the CBF constraints. 
\item In view of practical implementation, and since the convergence guarantees of the proposed algorithm are asymptotic, we propose a truncation mechanism for early termination. This comes at the cost of sacrificing strict guarantees of satisfying the CBF constraints, however, it reduces the communication and computation burden of an asymptotic algorithm. Moreover, it is accompanied with a verification scheme that provides probabilistic guarantees on safety constraint violation.
\item  The proposed verification scheme can be applied more generally to verify safety for multi-agent systems. In particular, instead of verifying safety over the whole state-space, which is challenging for multi-agent systems, we propose a scenario-based verification algorithm for a probabilistic quantification of safety by means of satisfying CBF constraints. A sequential sampling algorithm is proposed to sample scenarios efficiently in a distributed fashion. We accompany our solution with a probabilistic safety certificate; to achieve this, we extend the state-of-the-art result \cite[Theorem 1]{garatti2019risk} to a multi-agent setting. Both lower and upper bounds on the probability of {violating CBF constraints} are established, while the safety verification program is also shown to be amenable to parallelized computation.
\end{enumerate}
}

\subsection{Organization}
Section \ref{sec:dscl} proposes our distributed safe control design algorithm, including a truncated version and the associated mathematical analysis. Section \ref{sec:safetyverification} provides the distributed safety verification scheme, and the distributed scenario sampling algorithm. Section \ref{sec:simulation} demonstrates the control design and safety verification algorithms on a multi-robot system collision avoidance case study. Section \ref{sec:conc} concludes the paper and provides some directions for future research.

\section{Preliminaries}
\subsection{Notation}
\label{sec:nota}
We use $\mathbb{R}$, $\mathbb{R}^N$ to represent the set of one-dimensional, and $N$-dimensional real numbers, respectively.
A continuous function $\alpha(\cdot):(-b,a)\to (-\infty,+\infty)$ is said to be an extended class-$\mathcal{K}$ function for positive $a$ and $b$, if it is strictly increasing and $\alpha(0)=0$. $\mathcal{G}=(\mathcal{V},\mathcal{E})$ denotes a graph with a nodes set $\mathcal{V}$ and an edge set $\mathcal{E}$. Boldface symbols are used as stacked vectors for scalar or vector elements, e.g., $\boldsymbol{x}=[x_1^\top,\ldots,x_N^\top]^\top$. {For matrices $g_1,\ldots,g_N$, $\mathrm{diag}(g_1,\ldots,g_N)$ denotes the corresponding block diagonal matrix.} $I$ is an identity matrix, with its dimension being clear from the context. For a set $\mathcal{K}$, $|\mathcal{K}|$ denotes its cardinality. For a set $\mathcal{S}$, $\mathrm{Int}(\mathcal{S})$ denotes the interior.

\subsection{Control Barrier Functions}
Consider a nonlinear control-affine system
\begin{equation}\label{eq:nonlnsys}
  \dot x = f(x)+g(x)u,
\end{equation}
with $x(t) \in{ \mathcal{X}}\subset\mathbb{R}^n$, $u(t) \in\mathcal{U}\subset \mathbb{R}^m$, $f(x):{ \mathcal{X}}\to\mathbb{R}^n,$ and $g(x):{ \mathcal{X}}\to \mathbb{R}^{n\times m}$. Both $f$ and $g$ are further assumed to be {locally Lipschitz continuous on a compact set $\mathcal{X}\subset \mathbb{R}^n$.} { We denote by $x(u(\cdot),t,x_0)$ the state of the system at time $t$ starting from $x_0$, under a local Lipschitz continuous control law $u(\cdot)$.}

The safe set $\mathcal{S}$ is represented by the zero-super level set of a function $s(x)$. Dually, the unsafe set $\bar{\mathcal{S}}$ can be defined as the complementary set. With this formulation, the safe control design problem boils down to finding $u(\cdot)\in\mathcal{U}$, such that { $s(x(u(\cdot),t,x_0))\ge 0$} for any $t$. To achieve this, a control barrier function-based quadratic programming approach was proposed \cite{ames2016control}.

% Control barrier functions are an extension to barrier certificates \cite{prajna2004safety} for safety verification. It has been revealed in these papers that safety is closely related to the notion of \textit{control invariance}.
% \begin{defn}[\protect{\cite[Definition 3]{ames2016control}}]\label{def:invariance}
% A set $\mathcal{B} \subset \mathbb{R}^n$ is said to be control invariant with respect to \eqref{eq:nonlnsys}, if for any $x_0\in\mathcal{B}$, there exists $u\in\mathcal{U}$ such that ${x(u,t,x_0)}\in\mathcal{B}, \forall t\ge 0$.
% \end{defn}

% The relationship between safety and control invariance is demonstrated in the following equivalence lemma.
% \begin{lem}[\protect{\cite{taly2009deductive}}]\label{lem:safetyinvariance}
% System \eqref{eq:nonlnsys} is able to maintain safety under $\mathcal{S}$, if and only if there exists a control invariant set $\mathcal{B}\subseteq\mathcal{S}$. 
% \end{lem}
% Clearly, given a control invariant set $\mathcal{B}$, a safe control input $u(x)$ always exists for any $x\in\mathcal{B}$. The control barrier function approach answers the question of how to design a closed loop safe control input $u(x)$ inside $\mathcal{B}$. The notion of control barrier functions is related to the notion of extended class-$\mathcal{K}$ functions.
\begin{defn}
For the control-affine dynamical system \eqref{eq:nonlnsys}, a continuously differentiable function $b(\cdot):\mathbb{R}^n\to \mathbb{R}$ is said to be a control barrier function, if there exists an extended class-$\mathcal{K}$ function $\alpha(\cdot)$, such that for any $x\in\mathcal{B}$,
\begin{equation}\label{eq:cbf}
 \mathop{\sup\limits_{u\in\mathcal{U}}}[\mathcal{L}_fb(x)+\mathcal{L}_gb(x)u+\alpha(b(x))]\ge0.
\end{equation}
Here $\mathcal{L}_fb(x)$ and $\mathcal{L}_gb(x)$ are Lie derivatives, which are defined by $\mathcal{L}_fb(x):=\frac{\partial b(x)}{\partial x}f(x)$ and $\mathcal{L}_gb(x):=\frac{\partial b(x)}{\partial x}g(x)$, respectively.
\end{defn}
Given a control barrier function $b(x)$, the control admissible set corresponding to \eqref{eq:cbf} is defined by
\begin{equation}\label{eq:cbfset}
  K_{cbf}(x):=\{u\in\mathcal{U}:\mathcal{L}_fb(x)+\mathcal{L}_gb(x)u+\alpha(b(x))\ge0\}.
\end{equation}
\begin{thm}{\cite[Corollary 2]{ames2016control}}\label{th:theo1}
Consider a control barrier function $b(x)$. Then for any $x\in\mathcal{B}$, any locally Lipschitz continuous controller $u(x)$ such that $u(x)\in K_{cbf}(x)$ will render the set $\mathcal{B}$ forward invariant.
\end{thm}

{\subsection{Scenario Optimization}
Robust optimization offers a methodology to immunize decisions against uncertainty. 
An uncertain optimization problem is formulated as
\begin{equation}\label{eq:uncertain}
\begin{split}
  \min_{z\in\mathcal{Z}}~&c^\top z\\
  \mathrm{subject~to}~&z\in\mathcal{Z}_{x},~\text{for all}~x\in\mathcal{H},
\end{split} 
\end{equation}
where $z\in\mathbb{R}^n$ is a decision variable constrained by a set $\mathcal{Z}\subseteq\mathbb{R}^n$ and, $c\in\mathbb{R}^n$ is a constant vector. 
The uncertain constraint set $\mathcal{Z}_{x}$ is parameterized by an uncertain parameter $x$, which is a random variable defined on a probability space $(\mathcal{H},\mathcal{F},\mathbb{P})$. 
Even in the case where $\mathcal{Z}_{x}$ is convex for any $x \in \mathcal{H}$, 
if the uncertain parameters' domain $\mathcal{H}$ is continuous or even unknown, the robust optimization problem is usually hard (or even impossible) to solve. The so-called scenario approach, on the other hand, proposes to solve the problem over finite empirical records, named \emph{scenarios}, and accompany the resulting solution with probabilistic guarantees on its feasibility properties. 
The corresponding scenario optimization problem can be formulated as
\begin{equation}\label{eq:scenario}
\begin{split}
  \min_{z\in\mathcal{Z}}~&c^\top z\\
  \mathrm{subject~to}~&z\in\bigcap_{r=1,\ldots,R}\mathcal{Z}_{x^{(r)}},
\end{split} 
\end{equation}
where $x^{(r)}$, $r=1,\ldots,R$ are \emph{scenarios} sampled independently from the set $\mathcal{H}$. If $\mathcal{Z}_{x}$ is convex for any $x \in \mathcal{H}$, the scenario optimization \eqref{eq:scenario} is a convex optimization problem which can be solved efficiently. 

\begin{defn}[violation probability]\label{def:violation_probability}
  The violation probability of a a given $z\in\mathcal{Z}$ is defined as $V(z)=\mathbb{P}\{x\in\mathcal{H}:z\notin\mathcal{Z}_x\}$.
\end{defn}
Clearly, the optimal solution of \eqref{eq:scenario} satisfies $z^*\in\bigcap_{r=1,\ldots,R}\mathcal{Z}_{x^{(r)}}$, but is not necessarily within $\mathcal{Z}_x$ for an arbitrary new $x\in\mathcal{H}$. i.e., we do not necessarily have $V(z^*) = 0$. In fact, $z^*$ is itself a random variable as it depends on the choice of the scenarios $x^{(r)}$, $r = 1,\ldots,R$. To align with our subsequent developments, we will characterize $V(z^*)$ for a slightly more general scenario program; to this end, 
consider the following scenario optimization problem with relaxed constraints:
\begin{align}\label{eq:relax-scenario}
    \min_{z\in\mathcal{Z},\xi^{(r)}\ge 0,r=1,\ldots,R}&c^\top z +\rho\sum_{r=1}^R \xi^{(r)}\nonumber\\
    \mathrm{subject~to~}&h(z,x^{(r)})\le \xi^{(r)},r=1,\ldots,R,
\end{align}
where $x^{(r)},r=1,\ldots,R$ are independently sampled from $(\mathcal{H},\mathcal{F},\mathbb{P})$. Notice that here we consider the explicit characterization of the constraint set $\mathcal{Z}_{x^{(r)}}$ through functions $h(z,x^{(r)})$, $r = 1,\ldots,R$.

A constraint $z\in\mathcal{Z}_{x^{(r)}}$ is called a support constraint if its removal (while the other constraints are maintained) changes the solution $z^*$. We impose the following assumption.

\begin{assum}[\protect{\cite[Assumption 2]{garatti2019risk}}]\label{ass:degeneracy}
    Consider problem \eqref{eq:relax-scenario} and assume that a unique optimal solution $(z^*,\{\xi^{*,(r)}\}_{r=1}^R)$ exists almost surely with respect to the choice of $\{x^{(r)}\}_{r = 1}^R$. We further assume that the optimal solution $(z^*,\{\xi^{*,(r)}\}_{r=1}^R$ of \eqref{eq:relax-scenario} coincides almost surely with respect to the choice of the scenarios $x^{(r)}$, $r = 1,\ldots,R$ with the solution that is obtained after eliminating all the constraints that are not of support.
\end{assum}

The violation probability $V(z^*)=\mathbb{P}\{x\in\mathcal{H}:f(z^*,x)>0\}$ can be then characterized by the following theorem.
\begin{thm}[\protect{\cite[Theorem 4]{garatti2019risk}}]\label{th:scenario}
Consider the optimization problem \eqref{eq:relax-scenario}. Suppose that its optimal solution $(z^*,\{\xi^{*,(r)}\}_{r=1}^R$ satisfies Assumption \ref{ass:degeneracy}. Given a confidence parameter $\beta\in(0,1)$, for any $k=0,1,\ldots,R-1$ consider the polynomial equation in the $t$ variable
\begin{align}
&\left(\begin{array}{l}
{{ R}}\\
{k}
\end{array}\right)t^{{ R}-k}-\frac{\beta}{2{ R}}\sum_{j=k}^{{ R}-1}\left(\begin{array}{l}
{j}\\
{k}
\end{array}\right)t^{j-k}\nonumber\\
&-\frac{\beta}{6{ R}}\sum_{j={ R}+1}^{4{ R}}\left(\begin{array}{l}
{j}\\
{k}
\end{array}\right)t^{j-k}=0,\label{eq:pre-sc}
\end{align}
and for $k=R$ consider the polynomial equation
\begin{equation}\label{eq:pre-sc2}
    1-\frac{\beta}{6R}\sum_{i=R+1}^{4R}\left(\begin{array}{l}
{j}\\
{k}
\end{array}\right)t^{j-R}=0.
\end{equation}
For any $k=0,\ldots,R-1$, \eqref{eq:pre-sc} has exactly two solutions in $[0,+\infty)$, which we denote with $\underline{t}(k)$ and $\overline{t}(k)$ $(\underline{t}(k))\le \overline{t}(k)$. Instead, \eqref{eq:pre-sc2} has only one solution in $[0,+\infty)$, which we denote with $\overline{t}(R)$, while we define $\underline{t}(R)=0$. Let $\underline{\epsilon}(k):=\max\{0,1-\overline{t}(k)\}$ and $\overline{\epsilon}(k):=1-\underline{t}(k),k=0,1,\ldots,R$. We then have that
\begin{equation}
    \mathbb{P}^R\{\underline{\epsilon}({s}^*)\le V(z^*)\le \overline{\epsilon}({s}^*)\}\ge 1-\beta,
\end{equation}
where $s^*$ is the number of $x^{(r)}$'s for which $h(z^*,x^{(r)})\ge 0$.
\end{thm}
}

\section{Distributed Safe Control Law}
\label{sec:dscl}
% After proposing \eqref{eq:ldop} and its distributed solution, we are now in a position to use the proposed algorithm for \emph{distributed safe control input design}. 

Consider an $N$-agent system with the dynamics of the $i$-th agent described by
\begin{equation}\label{eq:dynamics}
  \dot x_i=f_i(x_i)+g_i(x_i)u_i,
\end{equation}
where $x_i(t)\in{ \mathcal{X}_i\subset}\mathbb{R}^{n_i}$ denotes its state, $u_i\in\mathcal{U}_i\subseteq \mathbb{R}^{m_i}$ denotes its control input, and $\mathcal{U}_i$ is a convex set. The dynamics $f_i(x_i):{ \mathcal{X}_i}\to\mathbb{R}^{n_i}$ and $g_i(x_i):{ \mathcal{X}_i}\to\mathbb{R}^{n_i}\times\mathbb{R}^{m_i}$ are both locally Lipschitz-continuous {on a compact set $\mathcal{X}_i\subset\mathbb{R}^{n_i}$, which represents the domain of each agent.} Vector $\boldsymbol{x}=[x_1^\top,\ldots,x_N^\top]^\top$ stacks the states of all systems, $\boldsymbol{u}=[u_1^\top,\ldots,u_N^\top]^\top$ stacks the control inputs, while {$f(\boldsymbol{x})=[f_1(x_1)^\top,\ldots,f_N(x_N)^\top]^\top$, $g(\boldsymbol{x})=\mathrm{diag}(g_1(x_1),\ldots,g_N(x_N))$} stack the dynamics for each agent. {The domain and control admissible set for the multi-agent system are then defined by
\begin{equation*}
  \mathcal{X}:=\prod_{i=1}^N\mathcal{X}_i, \quad \mathcal{U}:=\prod_{i=1}^N\mathcal{U}_i,
\end{equation*}
where $\prod$ represents the Cartesian product for the state space of all the agents. Given that all $\mathcal{X}_i$, $i=1,\ldots,N$, are assumed to be compact, compactness of $\mathcal{X}$ is assured using Tychonoff's theorem \cite{wright1994tychonoff}.} In this way, the system dynamics of the whole multi-agent system can be compactly modeled by $\dot{\boldsymbol{x}}=f(\boldsymbol{x})+g(\boldsymbol{x})\boldsymbol{u}$.

The networked system is described by an undirected and connected graph $\mathcal{G}$, with nodes set $\mathcal{V}=\{1,\ldots,N\}$, and edges set $\mathcal{E}$ such that $\{i,j\}\in\mathcal{E}$ if agent $j$ communicates with agent $i$. Agents are partitioned in $E$ sub-networks with specific safety requirement. For each subgraph $\mathcal{G}_e$, $e\in\{1,\ldots,E\}$, the set of agents is $\mathcal{V}_e \subseteq \mathcal{V}$. Let $\boldsymbol{x}_e=[x_i^\top]^\top_{i\in\mathcal{V}_e}$ be the stacked states in the $e$th sub-network. Each agent $i$ can communicate and cooperate with its neighbour $j\in\mathcal{N}_i$ to stay safe inside sub-network $e$ by ensuring 
\begin{equation}
\boldsymbol{x}_e(t)\in\mathcal{S}_e:=\{\boldsymbol{x}_e:s_e(\boldsymbol{x}_e)\ge 0\},~\forall t\ge 0,
\end{equation}
where $s_e(\cdot)\in\mathbb{R}$. {Define $\mathcal{S}:=\prod_{i=1}^E\mathcal{S}_e$, and let \(\mathcal{C}_i\) denote the set of indices representing the safety constraints associated with agent $i\in \{1,\ldots,N\}$. Specifically, for a given agent $i\in \{1,\ldots,N\}$, \(\mathcal{C}_i\) contains all indices \(e\) for which constraint \(e\) applies to agent \(i\). Given that each safety constraint involves a sub-network of agents, $\mathcal{C}_i$ also describes the set of indices of sub-networks that agent $i$ belongs to. As a result, agent $i$ belongs to sub-networks $\mathcal{G}_e, e\in\mathcal{C}_i$. Figure \ref{fig:networks} illustrates pictorially the relationship between $\mathcal{V}_e$ and $\mathcal{C}_i$.
\begin{figure}
    \centering
    \includegraphics[width=0.6\linewidth]{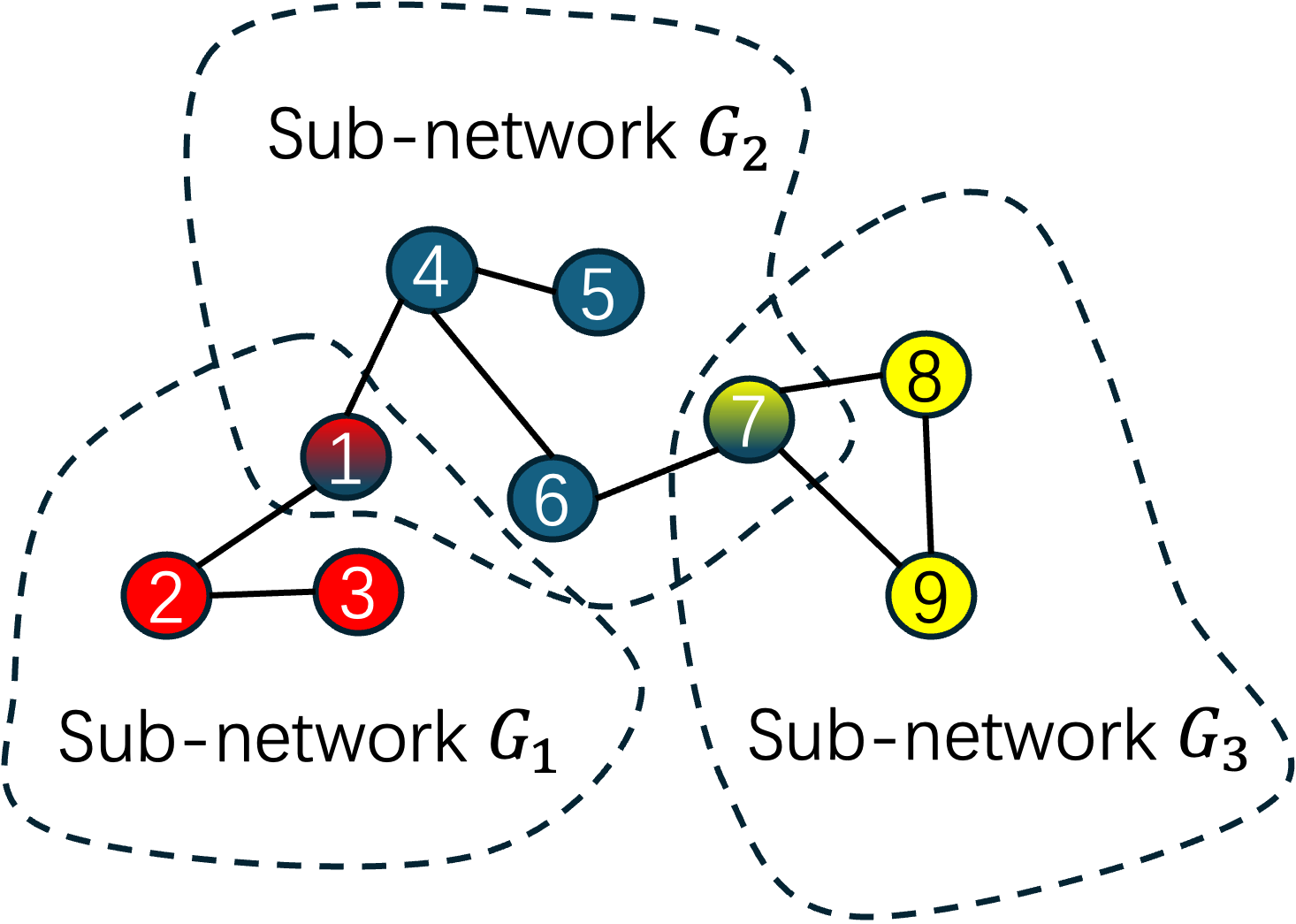}
    \caption{{ Pictorial illustration of a connected network $\mathcal{G}$ with 9 agents, where agents $1, 2,$ and $3$ form the sub-network $\mathcal{G}_1$ with safe set $\mathcal{S}_1$, agents $1, 4, 5, 6$ and $7$ form the sub-network $\mathcal{G}_2$ with a safe set $\mathcal{S}_2$, and agents $7, 8,$ and $9$ form the sub-network $\mathcal{G}_3$ with a safe set $\mathcal{S}_3$. The set of agents in each sub-network is given by $\mathcal{V}_1=\{1,2,3\}$, $\mathcal{V}_2=\{1,4,5,6,7\}$, and $\mathcal{V}_3=\{7,8,9\}$. It can be observed that agent $1$ belongs to two sub-networks, $\mathcal{G}_1$ and $\mathcal{G}_2$, and thus $\mathcal{C}_1=\{1,2\}$. Similarly, agent $2$ belongs only to $\mathcal{G}_1$, and agent $7$ belongs to both $\mathcal{G}_2$ and $\mathcal{G}_3$, giving $\mathcal{C}_2=\{1\}$ and $\mathcal{C}_7=\{2,3\}$.}}
    \label{fig:networks}
\end{figure}
}

\begin{assum}\label{ass:connectivity}
For each $e=1,\ldots,E$, sub-network $\mathcal{G}_e$ is connected and undirected.
\end{assum}

{Connectivity allows communication among agents in every sub-network $\mathcal{G}_e,e\in\{1,\ldots,E\}$. Agents in $\mathcal{G}_e$ are then able to cooperatively design a controller $\boldsymbol{u}_e(\boldsymbol{x})$ for safety, satisfying $s_e(\boldsymbol{x}_e)\ge 0,$ for all $e=1,\ldots,E.$}

\begin{assum}\label{ass:barrier}
Given sets $\mathcal{X}$ and $\mathcal{S}_e,e=1,\ldots,E$, we assume there exist control barrier functions $b_e(\cdot)$, such that $\mathcal{B}_e:=\{\boldsymbol{x}_e:b_e(\boldsymbol{x}_e)\ge 0\}\subseteq \mathcal{S}_e,e=1,\ldots,E$. Define { $\mathcal{B}:=\prod_{e=1}^E\mathcal{B}_e$, and $\mathcal{H}:=\mathcal{B}\cap\mathcal{X}$. We further assume that $\mathrm{Int}(\mathcal{H})\ne \emptyset$.}
\end{assum}

{Assumption \ref{ass:barrier} directly implies that $\mathrm{Int}(\mathcal{S})\ne \emptyset$ and $\mathrm{Int}(\mathcal{B})\ne \emptyset$. This is essential for using CBF methods to design safe controllers. However, checking emptiness of these sets is a challenging task. When $s_e(\boldsymbol{x}_e)$, $b_e(\boldsymbol{x}_e)$, $e=1,\ldots,E$, are polynomial functions, and $\mathcal{X}$ is defined by polynomial functions as well, emptiness can be checked via sum-of-squares programming. We refer the reader to \cite{parrilo2000structured} for further details.}

\begin{assum}\label{ass:control-sharing}
  Consider the multi-agent system \eqref{eq:dynamics} and CBFs $b_e(\boldsymbol{x}_e)$, $e=1,\ldots,E$, and class-$\mathcal{K}$ functions $\alpha_{ie}(\cdot),i=1,\ldots,N,e\in\mathcal{C}_i$. For every $\boldsymbol{x}\in{\mathcal{B}}$, we assume there exists a locally Lipschitz $\boldsymbol{u}=[u_1^\top\in\mathcal{U}_1,\ldots,u_N^\top\in\mathcal{U}_N]^\top\in\mathcal{U}$, such that for any $e\in\{1,\ldots,E\}$:
  \begin{equation}\label{eq:sum-cbf}
    \sum_{i\in\mathcal{V}_e}\left(\frac{\partial b_e}{\partial x_i}(f_i(x_i)+g_i(x_i)u_i)+\alpha_{ie}(b_e)\right)\ge 0.
  \end{equation}
\end{assum}

{The summation in \eqref{eq:sum-cbf} follows from applying the chain rule and considering the partial derivative of $b_e(\boldsymbol{x}_e)$ with respect to the state $x_i$ of every agent $i\in\mathcal{V}_e$.

Assumption \ref{ass:control-sharing} guarantees the existence of one controller $\boldsymbol{u}$ that satisfies all CBF constraints. This property is also known as the control sharing property \cite[Definition 2]{xu2018constrained}. CBFs that satisfy Assumption \ref{ass:barrier} and \ref{ass:control-sharing} can be designed using sum-of-squares programming \cite{schneeberger2023sos}.}

Following \cite[Theorem 3]{ames2016control}, safety constraints can be incorporated in the CBF-QP formulation given by
\begin{align}\label{eq:centralizedqp}
J^*=\min_{{u}_i\in\mathcal{U}}&\sum_{i=1}^N ||u_i-u_i^{\mathrm{des}}(x_i)||_2^2\nonumber\\
        \text{s.t.} &\sum_{i\in\mathcal{V}_e}\left(\frac{\partial b_{e}}{\partial x_i}(f_i(x_i)+g_i(x_i)u_i)+\alpha_{ie}(b_{e})\right)\ge 0,\nonumber\\
    &\forall e\in\{1,\ldots,E\},
\end{align}
where $\alpha_{ie}(\cdot)$'s are class-$\mathcal{K}$ functions, and hence also $\sum_{i\in\mathcal{V}_e}\alpha_{ie}(\cdot)$ is also a class-$\mathcal{K}$. $u_i^{\mathrm{des}}(x_i)$ is a nominal stabilizing control input. 

{The CBF constraints in \eqref{eq:centralizedqp} are defined on the control inputs for multiple agents. If every agent regards the variables of other agents as {stationary}, \eqref{eq:centralizedqp} decomposes to a family of problems, one for each $i=1,\ldots,N$,
\begin{equation}\label{eq:split-local}
\begin{split}
\min_{{u}_i\in\mathcal{U}_i}&||u_i-u_i^{\mathrm{des}}(x_i)||_2^2\\
\mathrm{s.t.}&\frac{\partial b_e}{\partial x_i}\left(f_i(x_i)+g_i(x_i)u_i\right)+\alpha_{ie}(b_e)\ge 0,\forall e\in\mathcal{C}_i.
\end{split}
\end{equation}
Under Assumptions \ref{ass:barrier} and \ref{ass:control-sharing}, \eqref{eq:centralizedqp} is guaranteed to be feasible, but feasibility of \eqref{eq:split-local} is not ensured for every $i\in\{1,\ldots,N\}$. In this work, we propose an improved distributed framework for solving \eqref{eq:centralizedqp} with guaranteed feasibility.
}

Let
\begin{equation}\label{eq:substi}
  \begin{split}
    J_i(u_i)&=||u_i-u_i^{\mathrm{des}}(x_i)||_2^2, \\
    h_{ie}(u_i)&=-\left(\frac{\partial b_{e}}{\partial x_i}(f_i(x_i)+g_i(x_i)u_i)+\alpha_{ie}(b_{e})\right).
  \end{split}
\end{equation}

{ We then have the following \emph{safety} results.
\begin{prop}[\protect{\cite[Proposition 1]{tan2022compatibility}}]\label{prop:safety}
Consider Assumptions \ref{ass:barrier}, \ref{ass:control-sharing}. Let $\boldsymbol{u}^*_{\mathrm{nom}}(\boldsymbol{x})$ be the optimal solution of \eqref{eq:centralizedqp}. Suppose $\boldsymbol{u}_{\mathrm{nom}}^*(\boldsymbol{x})$ is locally Lipschitz continuous for every $\boldsymbol{x}\in\mathcal{B}$, then the set $\mathcal{B}$ is forward invariant under the vector field $f(\boldsymbol{x})+g(\boldsymbol{x})\boldsymbol{u}_{\mathrm{nom}}^*(\boldsymbol{x})$.
\end{prop}
}

{\begin{rem}
    Local Lipchitz continuity of $\boldsymbol{u}_{\mathrm{nom}}^*(\boldsymbol{x})$ is important for forward invariance of $\mathcal{B}$ under the vector field $f(\boldsymbol{x})+g(\boldsymbol{x})\boldsymbol{u}_{\mathrm{nom}}^*(\boldsymbol{x})$. This can be guaranteed if the CBF constraints are linearly independent, and there are no input constraints. For more general cases, (strong) forward invariance can be guaranteed for a discontinuous vector field, under certain regularity conditions on the different CBFs. Interested readers are referred to \cite{usevitch2020strong,isaly2024feasibility}, and \cite[Section 9]{garg2024advances} for a comprehensive review. As this is tangential to the focus of our work, we will concentrate on the distributed implementation of the QP induced from multi-CBFs \eqref{eq:centralizedqp}. 
\end{rem}}

Notice that, even not shown explicitly, $h_{ie}(u_i)$ depends on $x_i,i\in\mathcal{V}_e$. We also highlight that \eqref{eq:centralizedqp} is parameterized in $\boldsymbol{x}$, which can be thought of as constant as for the optimization problem in \eqref{eq:centralizedqp} is concerned. {Under Assumptions \ref{ass:barrier}, \ref{ass:control-sharing}, problem \eqref{eq:centralizedqp} is always feasible for all $x\in\mathcal{B}$.}
To begin with our analysis, we propose a relaxed version of \eqref{eq:centralizedqp} {to guarantee feasibility of the local problems in the proposed distributed algorithm. This will be clarified in the sequel}.
\begin{align}\label{eq:relaxedqp}
H^*=&\min_{\boldsymbol{u}\in\mathcal{U},\boldsymbol{\rho}\ge 0}H(\boldsymbol{u},\boldsymbol{\rho})\nonumber\\
    :=&\sum_{i=1}^N\left\{J_i(u_i)+\sum_{e\in\mathcal{C}_i}(\rho^2_{ie}+M_{i}\rho_{ie})\right\}\nonumber\\
\mathrm{subject~to~}&\sum_{i\in\mathcal{V}_e}h_{ie}(u_i) \le{\sum_{i\in\mathcal{V}_e}}\rho_{ie}, {\forall e\in\{1,\ldots,E\}}.
\end{align}
Feasibility of problem \eqref{eq:relaxedqp} is clear, as the positive variable $\boldsymbol{\rho}$ relaxes the linear constraints. 

{In view of an optimality analysis, we further impose the following constraint qualification assumption.
\begin{assum}\label{ass:slater}
  For every $\boldsymbol{x}\in\mathcal{B}$, there exists $\boldsymbol{u}(\boldsymbol{x}){ \in\mathcal{U}}$, such that $\sum_{i\in\mathcal{V}_e}h_{ie}<0$ for all $e=1,\ldots,E$. 
\end{assum} 
Assumption \ref{ass:slater} ensures strong duality for the nominal problem \eqref{eq:centralizedqp}. As a result, there also exists $\boldsymbol{u}(\boldsymbol{x})\in\mathcal{U}$ and $\boldsymbol{\rho}=0$, such that $\sum_{i\in\mathcal{V}_e}h_{ie}<\sum_{i\in\mathcal{V}_e}\rho_{ie}$, for all $e\in\{1,\ldots,E\}$. This demonstrates strong duality for the relaxed problem \eqref{eq:relaxedqp}.
}

Optimality is analyzed in the following lemma. 

{

\begin{lem}\label{lem:newproperty}
Consider Assumptions \ref{ass:barrier}, \ref{ass:control-sharing} and \ref{ass:slater}.
Denote the minimizers of \eqref{eq:centralizedqp} and \eqref{eq:relaxedqp}, by $\boldsymbol{u}_{\mathrm{nom}}^*(\boldsymbol{x})$ and $\left(\boldsymbol{u}_{\mathrm{rel}}^*(\boldsymbol{x}),\boldsymbol{\rho}^*\right)$, respectively. Let $\tilde{\boldsymbol{\mu}}^*$ be an optimal dual variable associated with the CBF constraint in \eqref{eq:centralizedqp}. If
\begin{equation}\label{eq:lem2eq1}
  M_{i}\ge\tilde{\mu}_{e}^*,\forall i\in\mathcal{V}_e, \forall e\in\{1,\ldots,E\},
\end{equation}
where $\tilde{\mu}_e^*$ is the $e$-th element of $\tilde{\boldsymbol{\mu}}^*$, then $\boldsymbol{u}_{\mathrm{rel}}^*(\boldsymbol{x})=\boldsymbol{u}_{\mathrm{nom}}^*(\boldsymbol{x})$, $\boldsymbol{\rho}^*=0$, and $\tilde{\boldsymbol{\mu}}^*$ is also an optimal dual solution of \eqref{eq:relaxedqp}. 
\end{lem}

}
\begin{pf}
  See Appendix.
\end{pf}
{
Lemma \ref{lem:newproperty} establishes a lower bound for $M_i,i=1,\ldots,N$, under which the optimal primal-dual solution of \eqref{eq:relaxedqp} coincides with that of \eqref{eq:centralizedqp}. The lower bound is determined by the optimal dual solution $\tilde{\boldsymbol{\mu}}$  of the unrelaxed problem \eqref{eq:centralizedqp}. Under Assumption \ref{ass:slater}, $\tilde{\boldsymbol{\mu}}^*$ is also bounded following \cite[Lemma 1]{nedic2009approximate}. In practice, one can select a large enough $M_i,i=1,\ldots,N$ to satisfy \eqref{eq:lem2eq1}.
}

\subsection{Full Control Law}
We now design an algorithm to solve the centralized CBF-QP problem \eqref{eq:centralizedqp} in a distributed manner with guaranteed feasibility across iterations; see Algorithm \ref{al:dcbf}.
\begin{algorithm}[h]
 \caption{Distributed Safe Control Design Algorithm for agent $i$ at $x_i$}
  \hspace*{\algorithmicindent} \textbf{Initialization} Arbitrary $\lambda_{il}^0, \forall l\in\mathcal{N}_i\cap\mathcal{V}_e$, $\forall e\in\mathcal{C}_i$.\\
  \hspace*{\algorithmicindent} \textbf{Receive} $x_l$ for any $l\in\mathcal{N}_i\cap\mathcal{V}_e$,$e\in\mathcal{C}_i$\\
  \hspace*{\algorithmicindent} \textbf{Send} $x_i$ to any $l\in\mathcal{N}_i\cap\mathcal{V}_e$, for $e\in\mathcal{C}_i$.\\
 \hspace*{\algorithmicindent} \textbf{Output:} Optimal control input $u_i^*$\\
 \vspace{-3ex}
 \begin{algorithmic}[1]\label{al:dcbf}
 \WHILE{Not reaching convergence}
 \STATE \textbf{Receive} \label{step:firstcommunication} $\lambda_{il}^k$ from $\forall l\in\mathcal{N}_i\cap \mathcal{V}_e,\forall e\in\mathcal{C}_i$.
 \STATE \label{step:firstcomputation} \textbf{Solve} $((u_i^{k},\boldsymbol{\rho}_i^{k}),\boldsymbol{\mu}_i^{k})$ as a primal-dual solution of the following optimization problem
 \begin{equation}\label{eq:dcbf}
   \begin{split}
     \min_{u_i,\boldsymbol{\rho}_i}~&J_i(u_i) +\sum_{e\in\mathcal{C}_i}(\rho_{ie}^2+ M_{i}\rho_{ie})\\
     \mathrm{s.t.}~&u_i\in\mathcal{U}_i,\rho_{ie}\ge0,\\
     &h_{ie}(u_i)+\sum_{l\in\mathcal{N}_i\cap\mathcal{V}_e}(\lambda_{il}^k-\lambda_{li}^k)\le \rho_{ie},\forall e\in\mathcal{C}_i.
   \end{split}
 \end{equation}
 \STATE \textbf{Receive} \label{step:secondcommunication} $\mu_{le}^{k}$ from agent $l\in\mathcal{N}_i\cap\mathcal{V}_e$.
 \STATE \textbf{Update} \label{step:secondcomputation} $\lambda_{il}$ by  
 \begin{align}\label{eq:trunsubupdate}
   &\lambda_{il}^{k+1}=\lambda_{il}^k-\gamma^k(\mu_{ie}^{k}-\mu_{le}^{k}).
 \end{align}
 \ENDWHILE
 \end{algorithmic}
\end{algorithm}
Since $h_{ie}(u_i)$ also depends on $x_l$ for $l\in\mathcal{V}_e\backslash \{i\}$, an additional communication round at the beginning of the algorithm is designed. For all $i=1,\ldots,N$, and $e\in\mathcal{C}_i$, agent $i$ is to receive $x_l$ from agent $l\in\mathcal{N}_i\cap\mathcal{V}_e$. Within a finite number of communication rounds, agent $i$ can gather all the other agents' states in sub-networks $e\in\mathcal{C}_i$. Then, for any $e\in\mathcal{C}_i$, functions $h_{ie}(u_i)$ can be constructed as in \eqref{eq:substi}.

There are two main computation and two communication steps in the algorithm. At the first computation step (Step \ref{step:firstcomputation}), agent $i$ solves the optimization problem \eqref{eq:dcbf} to obtain the optimal primal-dual solution $((u_i^{k},\boldsymbol{\rho}^{k}_i),\boldsymbol{\mu}^{k}_i)$, where $\boldsymbol{\rho}_i$ includes relaxation variables denoted by $\rho_{ie}$ (penalized in the cost by $M_{i}$), and $\boldsymbol{\mu}_i$ includes the dual variables $\mu_{ie}$, for all $e\in\mathcal{C}_i$. In practice, $\mu_{ie}$ corresponds to the constraints allocated to agent $i$, i.e. $h_{ie}(x_i)+\sum_{l\in\mathcal{N}_i\cap\mathcal{V}_e}(\lambda_{il}^k-\lambda_{li}^k)\le \rho_{ie}$. Moreover, the constraints in the distributed problem \eqref{eq:dcbf} are relaxed by an additional non-negative relaxation variable $\rho_{ie}$. This guarantees the feasibility of the local optimization problem. However, this does not necessarily imply satisfaction of the CBF constraints in \eqref{eq:centralizedqp} by using $\boldsymbol{u}^{k+1}$.

The first computation step uses auxiliary variables $\lambda_{il}^k$ and $\lambda_{li}^k$. The difference $\lambda_{il}^k-\lambda_{li}^k$ constitutes estimates of the neighbouring terms $h_{le}(u_l)$. $\lambda_{il}^0$ is initialized arbitrarily. As we will show in Theorem \ref{th:speed}, the initialization will not influence convergence to the optimizer. Among all these variables, $\lambda_{le}^k$ for $l\in\mathcal{N}_i\cap\mathcal{V}_e$ are updated and stored by neighbours. They are available to agent $i$ via communication in Step \ref{step:firstcommunication}. 
The second computation step is to update the local auxiliary variables \eqref{step:secondcomputation}. Part of the dual variables used in the update are received from the neighbours at Step \ref{step:secondcommunication}. Here the update is a gradient-like procedure, with stepsize $\gamma^k>0$.

{
\begin{rem}
    Algorithm \ref{al:dcbf} capitalizes on the primal-decomposition algorithm in \cite[Algorithm RSDD]{notarnicola2019constraint}, however, with several key extensions. First, the relaxation penalty in the cost includes a new quadratic term. This renders the cost function strongly convex, allowing for superior convergence properties and ensuring uniqueness of the minimizer across iterations. Moreover, for every agent $i\in\{1,\ldots,N\}$, each CBF constraint $e\in\mathcal{C}_i$ is relaxed by an individual relaxation variable $\rho_{ie}$. On the contrary, \cite[Algorithm RSDD]{notarnicola2019constraint} uses one relaxation variable for all the constraints. Multiple relaxation variables enable stricter satisfaction of CBF constraints across iterations. This is especially important when a particular $\rho^k_{ie_1}$ is significantly larger than the other ones $\rho^k_{ie_2},e_2\in\mathcal{C}_i\backslash e_1$. It should also be noted that Algorithm \ref{al:dcbf} is applicable to the case where $\mathcal{G}$ is divided into several sub-networks $\mathcal{G}_e,e\in\{1,\ldots,E\}$,
    while \cite[Algorithm RSDD]{notarnicola2019constraint} only deals with a single network. This becomes of importance for multi-agent applications where safety constraints are typically defined on several sub-networks.
\end{rem}
}

Among different types of distributed optimization algorithms, primal-decomposition methods, firstly proposed by \cite[Algorithm RSDD]{notarnicola2019constraint} is selected here for its ability to guarantee almost-safety across iterations. This is realized by allocating the auxiliary variables $\boldsymbol{\lambda}$, while balancing the safety requirement to every agent. We say ``almost" here since additional relaxation variables are introduced in every local optimization problem for feasibility. In {applications that require high control frequency}, the algorithm may stop before reaching convergence. When the relaxation variables $\boldsymbol{\rho}^k=0$ for a given $k>0$, then for any $e\in\{1,\ldots,E\}$ we have that
\begin{equation*}
\begin{split}
  &\sum_{i\in\mathcal{V}_e}h_{ie}(u_i^k)=\sum_{i\in\mathcal{V}_e}\underbrace {\left\{h_{ie}(u_i^k)+\sum_{l\in\mathcal{N}_i\cap\mathcal{V}_e}(\lambda_{il}^k-\lambda_{li}^k)\right\}}_{ \le 0}\le 0,
\end{split}
\end{equation*}
which implies that the CBF constraints are satisfied with $\boldsymbol{u}^k$.
The next theorem gives the convergence result. 
\begin{thm}\label{th:speed}
Consider Assumptions \ref{ass:connectivity}, \ref{ass:barrier}, \ref{ass:control-sharing}, \ref{ass:slater}, and let $M_i\ge \mu_{e}$ for every $i=1,\ldots,N$, $e\in\mathcal{C}_i$. For every agent $i=1,\ldots,N$, and any bounded $\boldsymbol{\lambda}^0$,
\begin{itemize}
  \item [(a)] if $\mathcal{U}_i\subset\mathbb{R}^{m_i}.$ Choose the sequence $\{\gamma^k\}_{k\ge 0}$, with each $
  \gamma^k> 0$, and $\sum_{k=0}^\infty \gamma^k=\infty$, $\sum_{k=0}^\infty(\gamma^k)^2<\infty$. Then we have $\lim_{k\to\infty}H(\boldsymbol{u}^k,\boldsymbol{\rho}^k)-J^*\to 0$, and $\boldsymbol{u}^k$ converges to the primal optimal solution of \eqref{eq:centralizedqp}.
  \item [(b)] {if $ \mathcal{U}_i=\mathbb{R}^{m_i}$, and for every $e\in\{1,\ldots,E\}$, $\sum_{i\in\mathcal{V}_e}h_{ie}(u_i)$ are linearly independent in $\boldsymbol{u}$. Let the step size $\gamma^k=\gamma>0$ be a small constant.} $H(\boldsymbol{u}^k,\boldsymbol{\rho^k})$ converges to the optimal cost $J^*$ in \eqref{eq:centralizedqp} sublinearly, i.e. $H(\boldsymbol{u}^k,\boldsymbol{\rho}^k)-J^*\le \frac{2||\boldsymbol{\lambda^0}-\boldsymbol{\lambda^*}||_2^2}{\gamma k}$, and $\boldsymbol{u}^k$ converges to the primal optimal solution of \eqref{eq:centralizedqp}.
\end{itemize}
% \begin{subequations}\label{eq:parameterselection}
% \begin{align}
%   \sum_{e\in\mathcal{C}_i}(\frac{(\rho_{ie}^k)^2}{M_i}+\rho_{ie}^k)>\epsilon_i,\forall i=1,\ldots,N,\label{eq:para1}\\
%   \frac{P_i}{M_i}<\frac{1-\theta}{\theta}\epsilon_i,\forall i=1,\ldots,N,\label{eq:para2}
% \end{align}
% \end{subequations}
% then $0\le\rho_{\mathrm{sum}}^{k+1}<\rho_{\mathrm{sum}}^k$.
\end{thm}
\begin{pf}
  See Appendix.
\end{pf}

{Given that \eqref{eq:centralizedqp} is guaranteed to be feasible under Assumption \ref{ass:control-sharing}, the optimal controller designed by Algorithm \ref{al:dcbf} is guaranteed to satisfy all the CBF constraints. However, this does not necessarily hold for $\boldsymbol{u}^k(\boldsymbol{x})$ with arbitrary $k$, if $\boldsymbol{\rho}^k(\boldsymbol{x})\ne 0$. However, terminating the algorithm early, and considering $\boldsymbol{u}^k(\boldsymbol{x})$ at the time of termination has many benefits in terms of reducing computation and communication complexity. This motivates the analysis of a truncated algorithm as presented in the next section.}

\subsection{Truncated Control Law}
Algorithm \ref{al:dcbf} can be implemented in a distributed fashion with ensured safety and optimality properties, however, it may not be suitable for control tasks that require high control frequency, i.e. multi-robot system control, as its theoretical properties are established in an asymptotic manner. This motivates the use of a \emph{truncated algorithm}, Algorithm \ref{al:truncated}, where the algorithm terminates after a finite number of iterations, denoted by $\eta$.

\begin{algorithm}[h]
 \caption{Truncated Distributed Safe Control Design Algorithm for agent $i$}
  \hspace*{\algorithmicindent} \textbf{Initialization} Predefined $\lambda_{il}^0, \forall l\in\mathcal{N}_i\cap\mathcal{V}_e$, $\forall e\in\mathcal{C}_i$, truncated parameter $\eta\in\mathbb{N}$\\  \hspace*{\algorithmicindent} \textbf{Receive} $x_l$ for any $l\in\mathcal{N}_i\cap\mathcal{V}_e$,$e\in\mathcal{C}_i$\\
  \hspace*{\algorithmicindent} \textbf{Send} $x_i$ to any $l\in\mathcal{N}_i\cap\mathcal{V}_e,e\in\mathcal{C}_i$\\
 \hspace*{\algorithmicindent} \textbf{Output:} Optimal control input $u_i^*$\\
 \vspace{-3ex}
 \begin{algorithmic}[1]\label{al:truncated}
 \WHILE{$k\le \eta$}
 \STATE steps 2, 3, 4 in Algorithm \ref{al:dcbf}
 \STATE step 5 in Algorithm \ref{al:dcbf}
 \ENDWHILE
 \end{algorithmic}
\end{algorithm}

Algorithm \ref{al:truncated} is computationally more efficient compared to Algorithm \ref{al:dcbf}, at the cost of potentially violating the control barrier function constraints. The violations are reflected in the non-zero relation variables $\boldsymbol{\rho}^\eta{(\boldsymbol{x})}$. In general, it is challenging to provide an explicit bound for $\eta$, under which $\boldsymbol{\rho}^\eta{(\boldsymbol{x})}=0$, as the distributed algorithm converges asymptotically as per Theorem \ref{th:speed}. Moreover, $\boldsymbol{\rho}^\eta{(\boldsymbol{x})}$ depends on the state $\boldsymbol{x}\in\mathcal{H}:=\mathcal{X}\cap\mathcal{B}$, which parameterizes the optimization problem \eqref{eq:relaxedqp}. {To quantify safety of the multi-agent system \eqref{eq:dynamics} with $\boldsymbol{u}(\boldsymbol{x})=\boldsymbol{u}^\eta(\boldsymbol{x})$, we study the problem of safety verification by means of CBFs. This is established in the following section.}

\section{Distributed Safety Verification}
\label{sec:safetyverification}
%In the previous sections, two distributed safe control input design algorithms are presented with additional relaxations to guarantee feasibility of the optimisation problems across iterations. The algorithm may terminate before reaching the optimal control input $\boldsymbol{u}^*(\boldsymbol{x})$ which certainly renders the system safe. Moreover, if for some $\boldsymbol{x}$, the leveraged $\boldsymbol{u}(\boldsymbol{x})$ corresponds to a non-zero $\boldsymbol{\rho}^k$, the CBF constraints are violated. We point out here that constraint violation is not equivalent to becoming unsafe at the current state, but describes the \emph{risk} of becoming unsafe along the current trajectories by means of the CBFs. 

{In this section we show how to verify safety for a multi-agent system for any $\boldsymbol{x}\in\mathcal{H}$, using the truncated controller $\boldsymbol{u}^\eta(\boldsymbol{x})$ designed by Algorithm \ref{al:truncated}. The verification is conducted by checking the \emph{risk} of becoming unsafe along the current trajectories by means of CBFs. We would like to measure the violations of the CBF constraints for the multi-agent system \eqref{eq:dynamics}, under the control law $\boldsymbol{u}^\eta(\boldsymbol{x})$. However, this problem becomes challenging as $\mathrm{Int}(\mathcal{H})\ne \emptyset$, and one would need to verify a safety property for an uncountable number of points. Instead of verifying this for any $\boldsymbol{x}\in\mathcal{H}$, we propose to verify over \emph{finite} scenarios, i.e. samples of $\boldsymbol{x}$, from $\mathcal{H}$. Notice that 
the multi-agent system under consideration (see \eqref{eq:dynamics}) is deterministic; however, we draw scenarios as a discrete approximation of $\mathcal{H}$. 
The scenario approach \cite{garatti2019risk} then provides the theoretical foundation for quantifying the probability that the solution that satisfies our safety property for a finite number of scenarios, satisfies this property when it comes to yet another realization of  $\boldsymbol{x}\in\mathcal{H}$. Such a generalization property is in turn probabilistic, with a probability measure implicitly defined using the mechanism employed to draw scenarios (see Section \ref{subsec:5b}).}

{We note here the analysis conducted in this section can be applied to, but not limited to the controller designed using Algorithm \ref{al:truncated}. The only requirement for the verified controller $\boldsymbol{u}(\boldsymbol{x})$ is locally Lipschitz continuous, which is necessary for the solution of the multi-agent system to be unique. We also highlight that in this section a CBF is only regarded as a verification criterion but not necessarily as a control design principle.} 

%The rest of this section is organized in three parts: Section \ref{subsec:5a} proposes the scenario-based verification program, and compares them with the other programs; ii) Section \ref{subsec:5b} develops a distributed scenario sampling scheme, and gives a formal independency guarantee; iii) Section \ref{subsec:5c} shows the distributed implementation of the verification program, and gives probabilistic results. 

\subsection{Scenario Based Safety Verification}\label{subsec:5a}

Consider an $N$-agent system \eqref{eq:dynamics} and a safe invariant set $\mathcal{B}$. Our objective is to verify whether all the CBF constraints are satisfied for the multi-agent system \eqref{eq:dynamics} using $\boldsymbol{u}(\boldsymbol{x})$, for any $\boldsymbol{x}\in\mathcal{H}$. {A new set ${\mathcal{Z}_{\boldsymbol{x}}}$ is introduced to represent the satisfaction of all the CBF constraints.}

{
\begin{align}
    &\mathcal{Z}_{\boldsymbol{x}}:=\nonumber\\
    &\left\{\boldsymbol{z}:\sum_{i\in\mathcal{V}_e}h_{ie}(u_i(\boldsymbol{x}))\le \sum_{i\in\mathcal{V}_e}z_{ie},\forall e\in\{1,\ldots,E\}\right\}.
\end{align}
{Here, $\boldsymbol{z}:=[z_{ie}]$, $\forall e\in\{1,\ldots,E\},\forall i\in\mathcal{V}_e$.} Then, if $0\in{\mathcal{Z}_{\boldsymbol{x}}},\forall \boldsymbol{x}\in\mathcal{H}$, we conclude that all CBF constraints are satisfied using $\boldsymbol{u}(\boldsymbol{x})$, for any $\boldsymbol{x}\in\mathcal{H}$. With a slight abuse of notation, we define $\mathcal{Z}^i_{\boldsymbol{x}}$ as
\begin{align}
    &\mathcal{Z}^i_{\boldsymbol{x}}:=\left\{\boldsymbol{z}:\sum_{k\in\mathcal{V}_e}h_{ke}(u_k(\boldsymbol{x}))\le \sum_{k\in\mathcal{V}_e}z_{ke},\forall e\in\mathcal{C}_i\right\}
\end{align}
to represent the satisfaction of CBF constraints that involve agent $i$, for every $i\in\{1,\ldots,N\}$. {Here, $u_k(\boldsymbol{x})$ denotes the control input of agent $k$.} If $0\in{\mathcal{Z}_{\boldsymbol{x}}^i},~\forall \boldsymbol{x}\in\mathcal{H}$, the CBF constraints that involve agent $i$ are satisfied using $\boldsymbol{u}(\boldsymbol{x})$. Conversely, if $0\notin{\mathcal{Z}_{\boldsymbol{x}}^i}$, at least one CBF constraint that involves agent $i$ is violated, for some $\boldsymbol{x}\in\mathcal{H}$. Therefore, ${\mathcal{Z}_{\boldsymbol{x}}}$ can be expressed as
\begin{equation}\label{eq:constraints}
  \mathcal{Z}_{\boldsymbol{x}}=\bigcap_{i=1}^N\mathcal{Z}_{\boldsymbol{x}}^i.
\end{equation}
}

We propose a scenario-based safety verification program as follows.
\begin{align}\label{eq:dscenvarification}
    \min_{\boldsymbol{z}\ge 0,\boldsymbol{\zeta}\ge 0}~&\sum_{i=1}^N\sum_{e\in\mathcal{C}_i} \left(z_{ie}^2+{H_i}\sum_{r=1}^{{ R}}\zeta_{ie}^{(r)}\right)\nonumber\\
    \mathrm{s.t.}~
    &\sum_{i\in\mathcal{V}_e}h_{ie}(u_i(\boldsymbol{x}^{(r)}))\le \sum_{i\in\mathcal{V}_e}(z_{ie}+\zeta_{ie}^{(r)}),\nonumber\\
    &{\forall e\in\{1,\ldots,E\}, \forall r\in\{1,\ldots,R\}},
\end{align}
% \begin{equation}\label{eq:scenvarification}\tag{SC-Verification}
%     \begin{split}
%     \min_{\boldsymbol{\rho}\le 0,\boldsymbol{\zeta}\ge 0}~&\sum_{e=1}^E\left( \rho_e^2+M\sum_{r=1}^{{ R}}\zeta_e^{(r)}\right)\\
%     \mathrm{subject~to~}
%     &\sum_{l\in\mathcal{V}_e}h_{le}(u_l(\boldsymbol{x}^{(r)}))\le \rho_e+\zeta_e^{(r)},\\
%     &\forall e=1,\ldots,E, \forall r=1,\ldots,{ R},
%   \end{split}
% \end{equation}
where scenarios $\boldsymbol{x}^{(r)}\in{ \mathcal{H}}$ for any $r=1,\ldots,{ R}$ are extracted according to some probability distribution to be clarified in the sequel. 
Throughout the section $\bar X=\{\boldsymbol{x}^{(1)},\ldots,\boldsymbol{x}^{({ R})}\}$ denotes the set of scenarios, where $\boldsymbol{x}^{(r)}=[(x^{(r)}_1)^\top,\ldots,(x^{(r)}_N)^\top]^\top\in\mathbb{R}^{\sum_{i=1}^Nn_i}$, for $r=1,\ldots,{ R}$, {and $R$ is the number of scenarios.} Relaxation variables $\boldsymbol{\zeta}$ are introduced, while ${H_i}>0$ is a penalty coefficient for every $i\in\{1,\ldots,N\}$. {Let $(\boldsymbol{z}^*(\boldsymbol{x}),\boldsymbol{\zeta}^*(\boldsymbol{x}))$ denote the optimal solution of \eqref{eq:dscenvarification}. In the sequel, we drop the dependency of $\boldsymbol{x}$ for simplicity.}

{ Program \eqref{eq:dscenvarification} is a data-driven QP, where all the constraints are linear based on the samples. If for any scenario $\boldsymbol{x}^{(r)},r=1,\ldots,R,$ and the corresponding control input $\boldsymbol{u}(\boldsymbol{x})$, all the CBF constraints are satisfied, then $\boldsymbol{\zeta}^*=0$. Conversely, $\boldsymbol{\zeta}^*\ne0$ represents a CBF constraint violation, and indicates the risk of being unsafe by means of CBF, up to level $\boldsymbol{z}^*$. Following Definition \ref{def:violation_probability}, the violation probability for \eqref{eq:dscenvarification} is defined by
\begin{equation}\label{eq:violation-safety}
    V(\boldsymbol{z}):=\mathbb{P}\left\{\boldsymbol{x}\in\mathcal{H}:\boldsymbol{z}\notin{\mathcal{Z}_{\boldsymbol{x}}}\right\}.
\end{equation}
Then, $V(\boldsymbol{z}^*)=\mathbb{P}\left\{\boldsymbol{x}\in\mathcal{H}:\boldsymbol{z}^*\notin{\mathcal{Z}_{\boldsymbol{x}}}\right\}$ represents the probability that at least one CBF constraint is violated up to $\boldsymbol{z}^*$, for any $\boldsymbol{x}\in\mathcal{H}$.
Our goal is to \emph{distributedly} characterize the violation probability $V(\boldsymbol{z}^*)$ using a \emph{finite} number of scenarios, i.e. samples of $\boldsymbol{x}$ from $\mathcal{H}$.}

\subsection{Sampling the Scenarios}
\label{subsec:5b}
The scenarios are sampled independently from the set ${ \mathcal{H}}$. For sampling we define a probability density $\pi(\boldsymbol{x})$ associated with set ${ \mathcal{H}}$ that satisfies
  $\int_{{ \mathcal{H}}}\pi(\boldsymbol{x})d\boldsymbol{x}=1.$
One typical choice of $\pi(\boldsymbol{x})$ is to set it according to the density of the uniform distribution, i.e., $\pi(x)=\pi^{\mathrm{uni}}(\boldsymbol{x})=\frac{1}{\int_{ \mathcal{H}}d\boldsymbol{x}}$.

{The existence of $\pi^{\mathrm{uni}}(\boldsymbol{x})$ is assured as $\mathcal{H}$ is a non-empty and compact set, due to Assumption \ref{ass:barrier}.} Then, $\boldsymbol{x}$ can be sampled ${ R}$ times independently from the distribution $\pi^{\mathrm{uni}}(\boldsymbol{x})$. Note that the choice of the probability distribution does not affect the probabilistic results established in the sequel due to the distribution-free nature of the scenario approach \cite[Section 3.1]{garatti2019risk}. Although the uniform distribution here is well-defined, the set ${\mathcal{H}}$ is defined implicitly as the intersection of multiple sets. Sampling a point from the proposed uniform distribution is rather arduous in practice, and every agent may not have access to ${ \mathcal{H}}$. Here, we provide a sequential algorithm to sample scenarios $\boldsymbol{x}^{(r)}$, $r=1,\ldots,R$. 

\begin{algorithm}[h]
 \caption{Scenarios Sampling Algorithm}
  \hspace*{\algorithmicindent} \textbf{Initialization} { Set $\mathcal{H}=\mathcal{B}\cap\mathcal{X}$}, failed times $F=0$.\\
 \hspace*{\algorithmicindent} \textbf{Output:} Scenario $\boldsymbol{x}^{(r)}$.\\
 \vspace{-3ex}
 \begin{algorithmic}[1]\label{al:sampling}
 \STATE Sample $x_1^{(r)}$ from $\pi_1(\boldsymbol{x})$.
 \FOR{$i=2,\ldots,N$}
  \STATE Construct a {set} $\mathcal{H}_i=\bigcap_{e\in\mathcal{C}_i}\mathcal{H}_{ie}$ following \eqref{eq:fixedinvariant}.
 \IF {$\mathcal{H}_i=\emptyset$}
 \STATE $F \leftarrow F+1$.
 \STATE go to $i=i-F$ ($i=1$ is step $1$).
 \ENDIF
 \STATE {Sample $x_i^{(r)}$ from distribution $\pi_i=\frac{1}{\int_{\mathcal{X}_i}dx}$.}
 \WHILE{{$x_i^{(r)}\notin\mathcal{H}_i$}}
 \STATE {Sample $x_i^{(r)}$ from distribution $\pi_i$.}
 \ENDWHILE
 \ENDFOR
 
 \end{algorithmic}
\end{algorithm}
The algorithm constructs the densities from which samples are extracted sequentially for each agent. We first define the sets from which samples are extracted for agent $i$ with part of the states of agents in the same sub-network $\mathcal{G}_e$ fixed.
\begin{align}\label{eq:fixedinvariant}
  { \mathcal{H}}_{ie}&=\left\{ \begin{array}{l}
\mathcal{X}_i,~\mathrm{if}~\exists l\in\mathcal{V}_e,\mathrm{such}~\mathrm{that}~l> i\\
\{x_i\in{ \mathcal{X}_i}|b_{e}(x_i,\{x_l^{(r)}\})\ge0)\},~\mathrm{otherwise}
\end{array} \right.
\end{align}
Let ${\mathcal{H}}_i := \bigcap\limits_{e\in\mathcal{C}_i} { \mathcal{H}}_{i{e}}$. The parameters in \eqref{eq:fixedinvariant} can all be collected by local communication, since only states of agents in the same sub-network are required.

In {Step} $1$, the first scenario $x_1^{(r)}$ associated with Agent $1$ is sampled from distribution $\pi_1=\frac{1}{\int_{\mathcal{X}}dx}$, since now there are no other agents involved to restrict the set for Agent $1$. Then, the sampling-construction procedures repeat sequentially from Agent $2$ to Agent $N$. {For $i=2,\ldots,N$, before sampling the scenario $x_i^{(r)}$, we first check whether $\mathcal{H}_i$ is empty (Step 4).} If {$\mathcal{H}_i=\emptyset$} (Step 5), then we return to the sampling-construction of agent $i-F$, $F\ne 1$ to avoid a deadlock on step $i$. The deadlock happens when for given scenarios $x_1^{(r)},\ldots,x_{i-2}^{(r)}$, the set ${ \mathcal{H}}_{i-1}$ is such that for any $ x_{i-1}^{(r)}\in{ \mathcal{H}}_{i-1}$, ${ \mathcal{H}}_i=\emptyset$. It is guaranteed that $F\le i-1$ for $i\ge 2$, since ${ \mathcal{H}}_1=\mathcal{X}_1\ne\emptyset$. {After finding feasible scenarios $x_1^{(r)},\ldots,x_{i-1}^{(r)}$, we sample the scenario $x_i^{(r)}$ for the $i$th agent from the uniform distribution $\pi_i$ (Step 8). The sampled scenario is then checked at Step 9. If $x_i^{(r)}\notin\mathcal{H}_i$, it will be sampled again following $\pi_1$. The loop will terminate in finite time since $\mathrm{Int}(\mathcal{H}_i\cap\mathcal{X})\ne \emptyset~\forall i\in\{1,\ldots,N\}$.}

\begin{prop}\label{pro:independency}
{Consider Assumptions \ref{ass:connectivity}, \ref{ass:barrier}, and assume scenarios $\boldsymbol{x}^{(r)}$, $r=1,\ldots,{ R}$ are sampled using Algorithm \ref{al:sampling}.} We then have that $\boldsymbol{x}^{(r)}\in{ \mathcal{H}}$, for all $r=1,\ldots,{ R}$. Moreover, all scenarios are independently and {identically} sampled.
\end{prop}
\begin{pf}
The feasibility result holds directly from the definition of every set ${ \mathcal{H}}_i$ in \eqref{eq:fixedinvariant} that $x_i^{(r)}$ is sampled from. As a result, we have $b_{ie}(x_i^{(r)},\{x_k^{(r)}\})\ge 0$ for any $i=1,\ldots,N$, $e\in\mathcal{C}_i$, and $k\in\mathcal{V}_e\backslash i$. Therefore, $\boldsymbol{x}^{(r)}\in{ \mathcal{H}}$. Moreover, for all $r=1,\ldots,{ R}$, $\boldsymbol{x}^{(r)}$ are independent since $x_1^{(r)}$, $r=1,\ldots,{ R}$ are independently sampled from the distribution $\pi_1$.

{At Step 6, when $F=i-1$, it returns Step 1 to resample $x_1^{(r)}$. This happens when there exists $e\in\mathcal{C}_2$, and $b_{e}$ is defined only on Agent $1$ and $2$, such that $\mathcal{H}_{2e}=\{x_2\in\mathcal{X}_2|b_{e}(x_2,x_1^{(r)})\ge 0\}=\emptyset$. $x_1^{(r)}$ will then be resampled from the distribution $\pi_1$ to make $\mathcal{H}_{2e}\ne \emptyset$. Therefore, the actual distribution $\tilde{\pi}_1$ from which $x_1^{(r)}$ is sampled is defined on a set $\tilde{\mathcal{X}_1}\subseteq\mathcal{X}_1$, which satisfies
\begin{equation}
    \{x_2\in\mathcal{X}_2:b_{e}(x_2,x_1^*)\ge 0\}\ne \emptyset,\forall x_1^*\in\tilde{\mathcal{X}_1}.
\end{equation}
It trivially holds that $\mathrm{Int}(\tilde{\mathcal{X}_1})\ne\emptyset$ since $\mathrm{Int}(\mathcal{H})\ne\emptyset$, from Assumption \ref{ass:barrier}.
$\tilde{\pi}_1$ can be different from $\pi_1$, but is identical for every $r=1,\ldots,R$. Similarly, the resampling mechanism implicitly defines distributions $\tilde{\pi}_2,\ldots,\tilde{\pi}_N$ that may be different from $\pi_2,\ldots,\pi_N$. But these distributions are identical for scenarios $\boldsymbol{x}^{(r)},r=1,\ldots,R$.}
\end{pf}

We note here that the elements in $\boldsymbol{x}^{(r)}$ are correlated, but this will not influence the independence results in Proposition \ref{pro:independency} since we seek independence across $r$.

\subsection{Distributed Safety Verification}
\label{subsec:5c}
%After sampling every scenario $\boldsymbol{x}^{(r)}$, the scenario represents the states to the networked system \eqref{eq:dynamics}. Agents then design their control law $\boldsymbol{u}(\boldsymbol{x}^{(r)})$ with their methods. As we have stated in the beginning of this section, the control law $\boldsymbol{u}(\boldsymbol{x})$ is not necessarily designed via solving the CBF-QP \eqref{eq:centralizedqp}, but can also be synthesized by other methods. The only requirement is that every agent has access to its control input. Without additional communication, $x_i^{(r)}$ and $u_i(\boldsymbol{x}^{(r)})$ are private resources to agent $i$. Although $u_i(\boldsymbol{x}^{(r)})$ may be designed on the scenarios (states) of other agents, in the verification step we still assume the scenarios are private since the design of control input is not \textit{a priori}. 

% The verification program \eqref{eq:scenvarification} is well-defined, however it can not be directly solved locally with Algorithm \ref{al:dcbf} since the problem is still coupled across agents as the decision variables $\rho_e$ is common variable for all agents in the sub-network $\mathcal{G}_e$. A commonly used approach to split the problem is to decompose the ``edge" variables into several ``node" variables, entailing additional consensus constraints. Following this idea, a decoupled version is proposed as
After sampling scenarios $\boldsymbol{x}^{(r)}$, $r=1,\ldots,{ R}$ using Algorithm \ref{al:sampling}, we are at the stage of solving the safety verification program \eqref{eq:dscenvarification}.

Letting the local cost function $J_i(\boldsymbol{z}_i,\boldsymbol{\zeta}_i)$, and constraint function $\hat{h}_{ie}(\boldsymbol{z}_i,\boldsymbol{\zeta}_i)$ be
\begin{align}\label{eq:scenariodistributed}
  J_i(\boldsymbol{z}_i,\boldsymbol{\zeta}_i)&=\sum_{e\in\mathcal{C}_i} \left(z_{ie}^2+{H_i}\sum_{r=1}^{{ R}}\zeta_{ie}^{(r)}\right),\nonumber\\
    \hat{h}_{ie}^{(r)}(\boldsymbol{z}_i,\boldsymbol{\zeta}_i)&={h}_{ie}(u_i(\boldsymbol{x}^{(r)}))-z_{ie}-\zeta_{ie}^{(r)},r=1,\ldots,{ R},
\end{align}
Algorithm \ref{al:dcbf} can be applied to solve the distributed scenario optimization problem \eqref{eq:dscenvarification}. The relaxation variables in Algorithm \ref{al:dcbf} are unnecessary, since every optimization sub-problem across iterations is solvable. We then have the following theorem as the main result on distributed probabilistic safety. {The following theorem constitutes the multi-agent counterpart of Theorem \ref{th:scenario}. Using the density functions constructed in Algorithm \ref{al:sampling} and considering Assumption \ref{ass:barrier}, there will be no repeated scenarios for $r=1,\ldots,R$. Therefore, eliminating all the constraints that are not in the support set for \eqref{eq:dscenvarification} will not change the optimal solution $\boldsymbol{z}^*$, and hence due to Assumption \ref{ass:barrier}, the non-degeneracy requirement of Assumption \ref{ass:degeneracy} is satisfied.}

{
\begin{thm}\label{th:probability} Let Assumptions \ref{ass:connectivity} and \ref{ass:barrier} hold. Consider the optimization problem \eqref{eq:dscenvarification} and let $(\boldsymbol{z}^*,\{\boldsymbol{\zeta}^{*,(r)}\}_{r=1}^R)$ be the optimal solution. Choose $\beta_i \in (0,1),i=1,\ldots,N$, and set $\beta=\sum_{i=1}^N \beta_i$. For $i=1,\ldots,N$, and $0\le k_i\le { R}-1$, consider the polynomial equation in the $t_i$ variable
\begin{equation}\label{eq:scenariopolynomial}
\begin{split}
    &\left(\begin{array}{l}
{{ R}}\\
{k_i}
\end{array}\right)t_i^{{ R}-k_i}-\frac{\beta_i}{2{ R}}\sum_{j=k_i}^{{ R}-1}\left(\begin{array}{l}
{j}\\
{k_i}
\end{array}\right)t_i^{j-k_i}\\
&-\frac{\beta_i}{6{ R}}\sum_{j={ R}+1}^{4{ R}}\left(\begin{array}{l}
{j}\\
{k_i}
\end{array}\right)t_i^{j-k_i}=0,
\end{split}
\end{equation}
while for $k_i={ R}$ consider the polynomial equation
\begin{equation}\label{eq:scenariopolynomial2}
  1-\frac{\beta}{6N}\sum_{j={ R}+1}^{4{ R}}\left(\begin{array}{l}
{j}\\
{k_i}
\end{array}\right)t_i^{j-{ R}}=0.
\end{equation}
For every $i=1,\ldots,N$ and any $k_i=0,\ldots,R-1$, Equation \eqref{eq:scenariopolynomial} has exactly two solutions in $[0,+\infty)$ denoted by $\underline t_i(k_i)$ and $\bar t_i(k_i)$, where $\underline t_i(k_i)\le \bar t_i(k_i)$. Instead, Equation \eqref{eq:scenariopolynomial2} has only one solution in $[0,+\infty)$, which we denote with $\overline{t}_i(R)$, while we define $\underline{t}_i(R)=0$. Let $\underline\epsilon_i(k_i):=\max\{0,1-\bar t_i(k_i)\}$, $\bar \epsilon_i(k_i):=1-\underline t_i(k_i)$, and $\underline \epsilon(s^*)=\sum_{i=1}^N\underline\epsilon_i(s_i^*)$, $\bar \epsilon(s^*)=\min\{\sum_{i=1}^N\bar\epsilon_i(s_i^*),1\}$. We then have that
\begin{equation}\label{eq:violation}
  \mathbb{P}^{{ R}}\left\{\frac{\underline\epsilon(s^*)}{N}\le V(\boldsymbol{z}^*)\le\bar\epsilon(s^*)
  \right\}\ge 1-\beta,
\end{equation}
where $s_i^*$ is the number of $\boldsymbol{x}^{(r)}$'s for which there exists $e\in\mathcal{C}_i$, such that $\sum_{k\in\mathcal{V}_e}h_{ke}(u_k(\boldsymbol{x}^{(r)}))\ge \sum_{k\in\mathcal{V}_e}z_{ke}^*$. Recalling Equation \eqref{eq:violation-safety}, the violation probability $V(\boldsymbol{z}^*)$ is defined by $V(\boldsymbol{z}^*)=\mathbb{P}\{\boldsymbol{x}\in\mathcal{H}:\boldsymbol{z}^*\notin{\mathcal{Z}_{\boldsymbol{x}}}\}.$
\end{thm}
}
\begin{pf}
  See Appendix.
\end{pf}

Theorem \ref{th:probability} is a generalization of \cite[Theorem 2]{garatti2019risk} to a multi-agent setting. It also extends \cite{margellos2017distributed} by determining the lower bound $\frac{\underline\epsilon(s^*)}{N}$.
Theorem \ref{th:probability} states that with confidence at least $1-\beta$, {the probability that the CBF constraints of the multi-agent system are violated by more than $\boldsymbol{z}^*$}, lies within the interval $[\frac{\underline\epsilon(s^*)}{N}$, $\bar \epsilon(s^*)]$.

\section{Simulation Results}
\label{sec:simulation}
The distributed safe control input design and safety verification algorithms are numerically validated on a multi-robot positions swapping problem. To facilitate comparison, we adopt a similar setup as in \cite{wang2017safety}. 
\subsection{Multi-Robot Position Swapping}
Robots are assigned different initial positions and are required to navigate towards target locations. In a distributed framework, robots are equipped with sensing and communication modules for collision detection and information sharing. A {network} of ten robots, indexed by $i=1,\ldots,10$ are considered, with double integrator dynamics
\begin{equation}\label{eq:robotdynamics}
  \begin{bmatrix}
  \dot{\boldsymbol{p}_i}\\\dot{\boldsymbol{v}_i}
  \end{bmatrix}=
  \begin{bmatrix}
  0&I_{2\times 2}\\
  0&0
  \end{bmatrix}
  \begin{bmatrix}
  \boldsymbol{p}_i\\
  \boldsymbol{v}_i
  \end{bmatrix}+
  \begin{bmatrix}
  0\\
  I_{2\times 2}
  \end{bmatrix}\boldsymbol{a}_i,
\end{equation}
where $\boldsymbol{p}_i\in\mathbb{R}^2$, $\boldsymbol{v}_i\in\mathbb{R}^2$ represent positions and velocities, and $\boldsymbol{a}_i\in\mathbb{R}^2$ is the control input, representing accelerations. The acceleration is limited as $||\boldsymbol{a}_i||_\infty\le a_i^{\max}$. $a_i^{\max}$ will be cleared in the sequel. Each robot is regarded as a disk centered at $\boldsymbol{p}_i$ with radius $D_i\in\mathbb{R}_+$. The safety certificate $s_{ij}(\boldsymbol{p},\boldsymbol{v})$ for collision avoidance between robot $i$ and $j$ is defined by
\begin{equation}\label{eq:robotsafety}
  s_{ij}(\boldsymbol{p},\boldsymbol{v}) = ||\Delta\boldsymbol{p}_{ij}||_2^2-D_{ij},
\end{equation}
where $\Delta\boldsymbol{p}_{ij}=\boldsymbol{p}_i-\boldsymbol{p}_j$, $D_{ij}=D_i+D_j$.
Note here that the system is heterogeneous as different robots have different mobility. Following \cite{wang2017safety}, the control barrier function for invariance certificates is then defined pair-wisely, as
\begin{align}
  b_{ij}(\boldsymbol{p},\boldsymbol{v}) &= \sqrt{2(a_i^{\max}+a_j^{\max})(||\Delta\boldsymbol{p}_{ij}||_2^2-D_{ij})}\nonumber \\&+\frac{\Delta\boldsymbol{p}_{ij}^\top}{||\Delta\boldsymbol{p}_{ij}||_2^2}\Delta\boldsymbol{v}_{ij}, \label{eq:robotbarrierfunction}
\end{align}
where $\Delta\boldsymbol{v}_{ij}=\boldsymbol{v}_i-\boldsymbol{v}_j$. The function $b_{ij}(\boldsymbol{p},\boldsymbol{v})$ is guaranteed to be a CBF since when $b_{ij}(\boldsymbol{p},\boldsymbol{v})>0$, collision can be avoided with maximum braking acceleration $\boldsymbol{a}^{\max}_i+\boldsymbol{a}^{\max}_j$ applied to robots $i$ and $j$. For $i=1,\ldots,5$, $\boldsymbol{a}_i^{\max}=1$, while for $i=6,\ldots,10$, $\boldsymbol{a}_i^{\max}=10$. Note that although $b_{ij}(\boldsymbol{p},\boldsymbol{v})$ is guaranteed to be a CBF for safety certificate $s_{ij}(\boldsymbol{p},\boldsymbol{v})$, the corresponding invariant set $\mathcal{B}=\prod\limits_{\{i,j\}\in\mathcal{E}}\mathcal{B}_{ij}$ is possibly empty. Intuitively, this is since robots cannot utilize the maximum braking force to avoid collision with multiple other robots simultaneously. This problem is beyond the scope of this paper, and we still adopt the CBF as in \eqref{eq:robotbarrierfunction}.

\begin{figure}[t]
   \centering

   \begin{subfigure}[b]{0.23\textwidth}
     \centering
     \includegraphics[width=\textwidth]{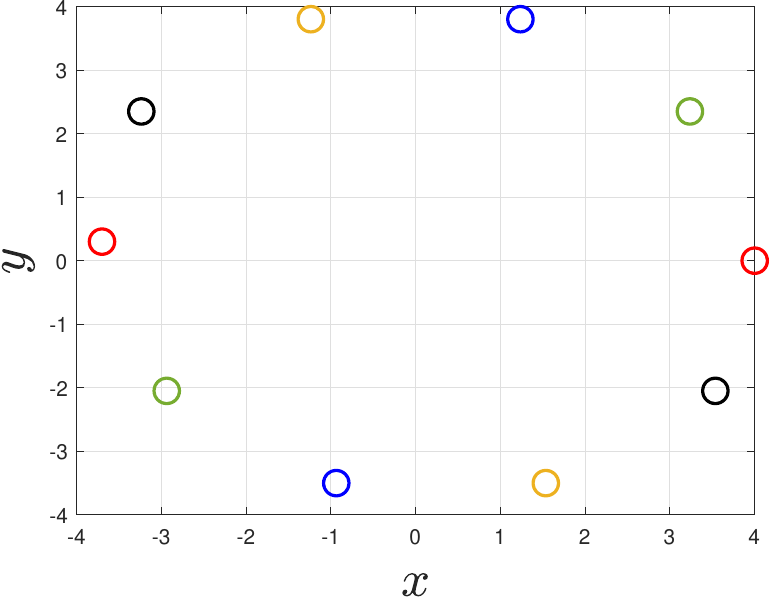}
     \caption{{ Time} 1}
   \end{subfigure}
   \hfill
   \begin{subfigure}[b]{0.23\textwidth}
     \centering
     \includegraphics[width=\textwidth]{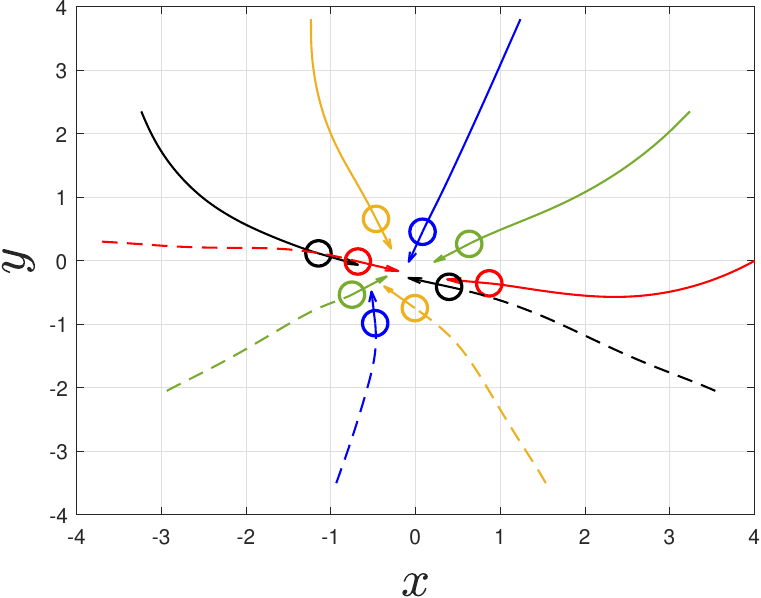}
     \caption{{ Time} 200}
   \end{subfigure}
   \hfill
   \begin{subfigure}[b]{0.23\textwidth}
     \centering
     \includegraphics[width=\textwidth]{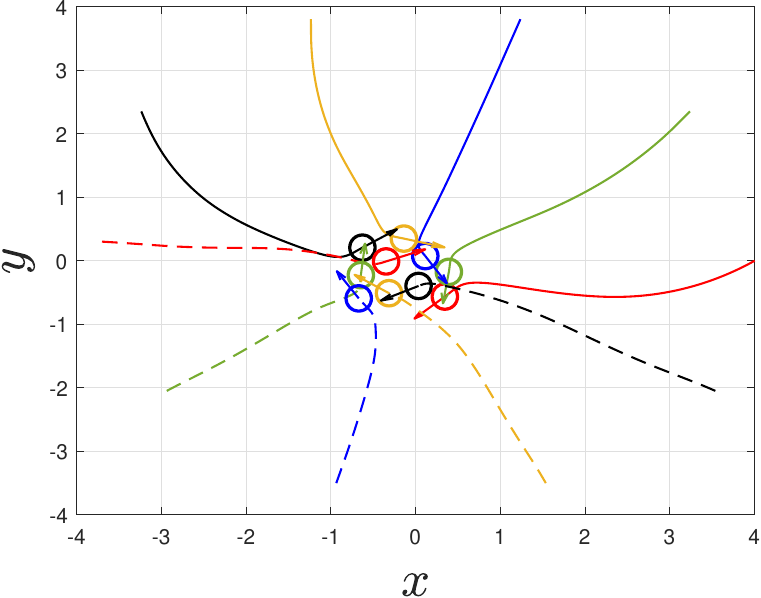}
     \caption{{ Time} 300}
   \end{subfigure}
     \hfill
   \begin{subfigure}[b]{0.23\textwidth}
     \centering
     \includegraphics[width=\textwidth]{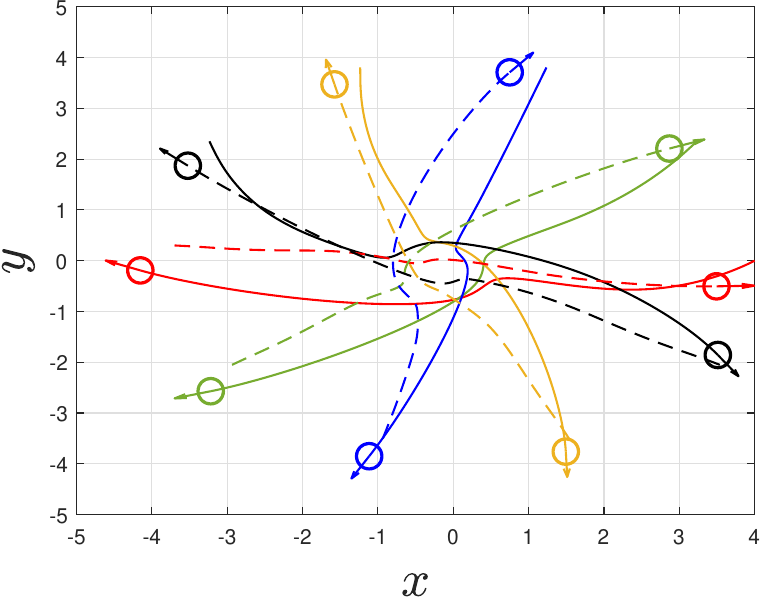}
     \caption{{ Time} 500}
   \end{subfigure}
     \caption{Trajectory of ten robots swapping positions according to Algorithm \ref{al:dcbf}. Robots with the same color are swapping positions, and avoiding collision with the others.}
     \label{fig:fullcontrol}
\end{figure}

\subsection{Distributed Control: Asymptotic Algorithm}
The distributed safe control design procedure of Algorithm \ref{al:dcbf} that exhibits asymptotic convergence and optimality guarantees that it is implemented for robots to swap positions with the opposite robots while avoiding collision. The resulting simulation results are shown in Figure \ref{fig:fullcontrol}.

\subsection{Distributed Control: Truncated Algorithm}
The truncated Algorithm \ref{al:truncated} is then implemented for the same setting, the truncation parameter $\eta=30$.

\begin{figure}[ht]
   \centering

   \begin{subfigure}[b]{0.23\textwidth}
     \centering
     \includegraphics[width=\textwidth]{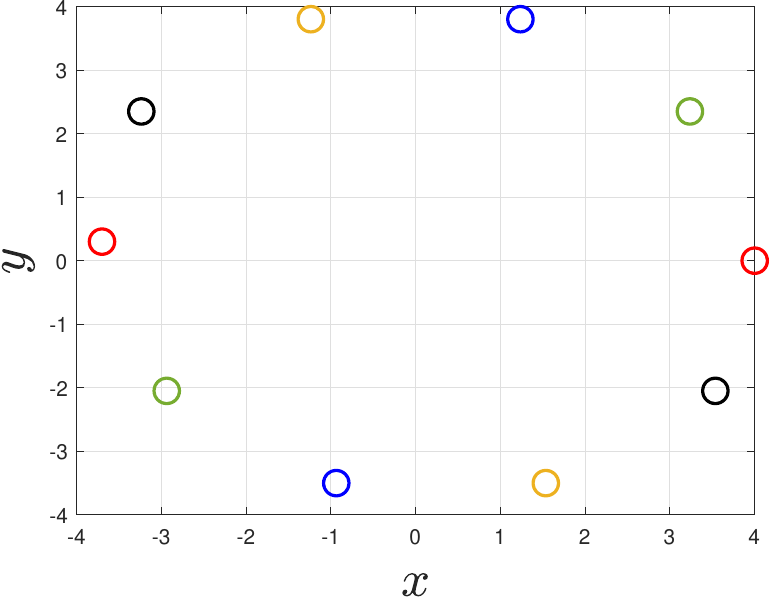}
     \caption{{ Time} 1}
   \end{subfigure}
   \hfill
   \begin{subfigure}[b]{0.23\textwidth}
     \centering
     \includegraphics[width=\textwidth]{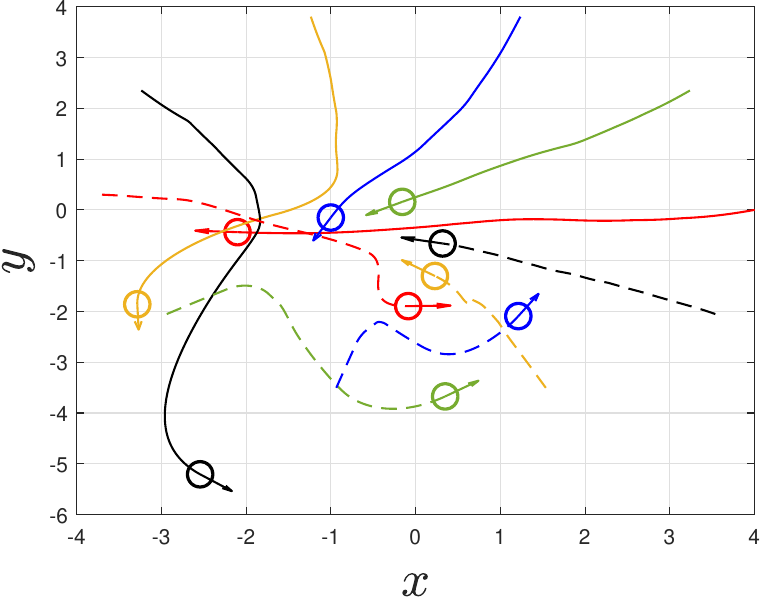}
     \caption{{ Time} 200}
   \end{subfigure}
   \hfill
   \begin{subfigure}[b]{0.23\textwidth}
     \centering
     \includegraphics[width=\textwidth]{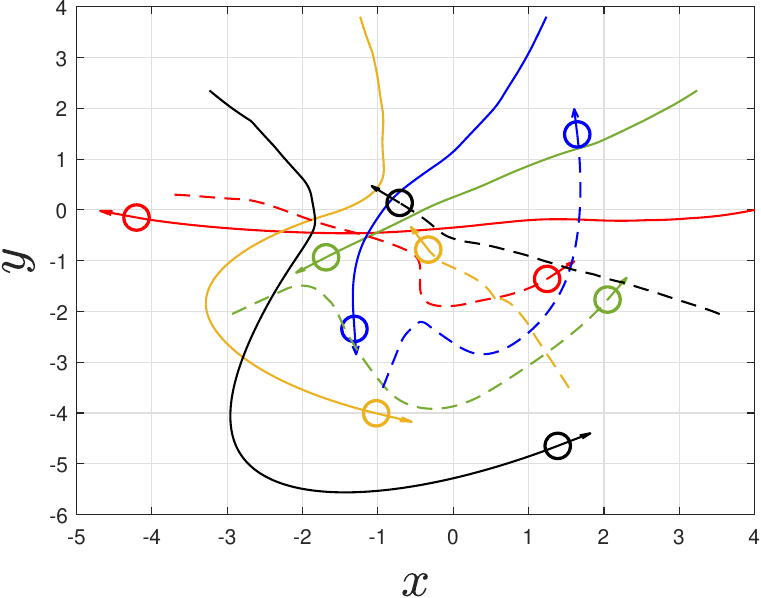}
     \caption{{ Time} 300}
   \end{subfigure}
     \hfill
   \begin{subfigure}[b]{0.23\textwidth}
     \centering
     \includegraphics[width=\textwidth]{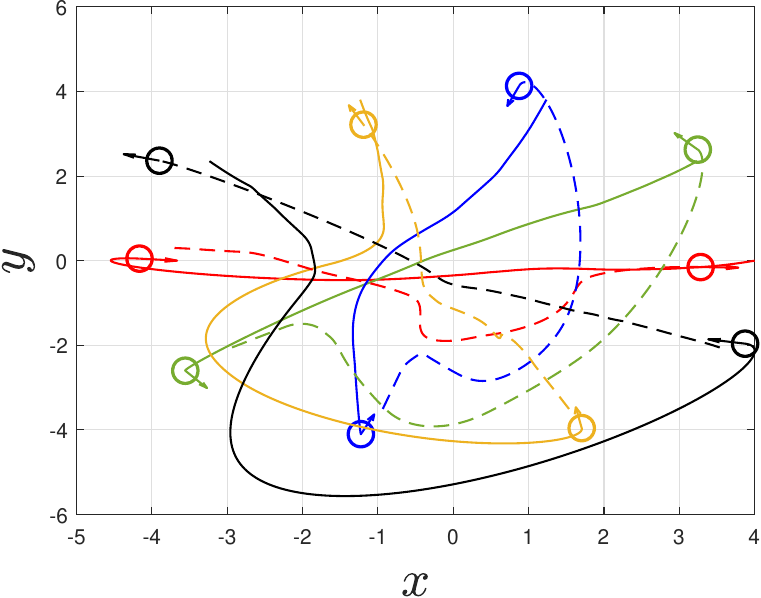}
     \caption{{ Time} 500}
   \end{subfigure}
     \caption{Trajectory of ten robots swapping positions while avoiding collision by means of Algorithm \ref{al:truncated}, with $\eta=30$.}
     \label{fig:truncontrol}
\end{figure}
\begin{figure}[h]
   \centering

   \begin{subfigure}[t]{0.23\textwidth}
     \centering
     \includegraphics[width=\textwidth]{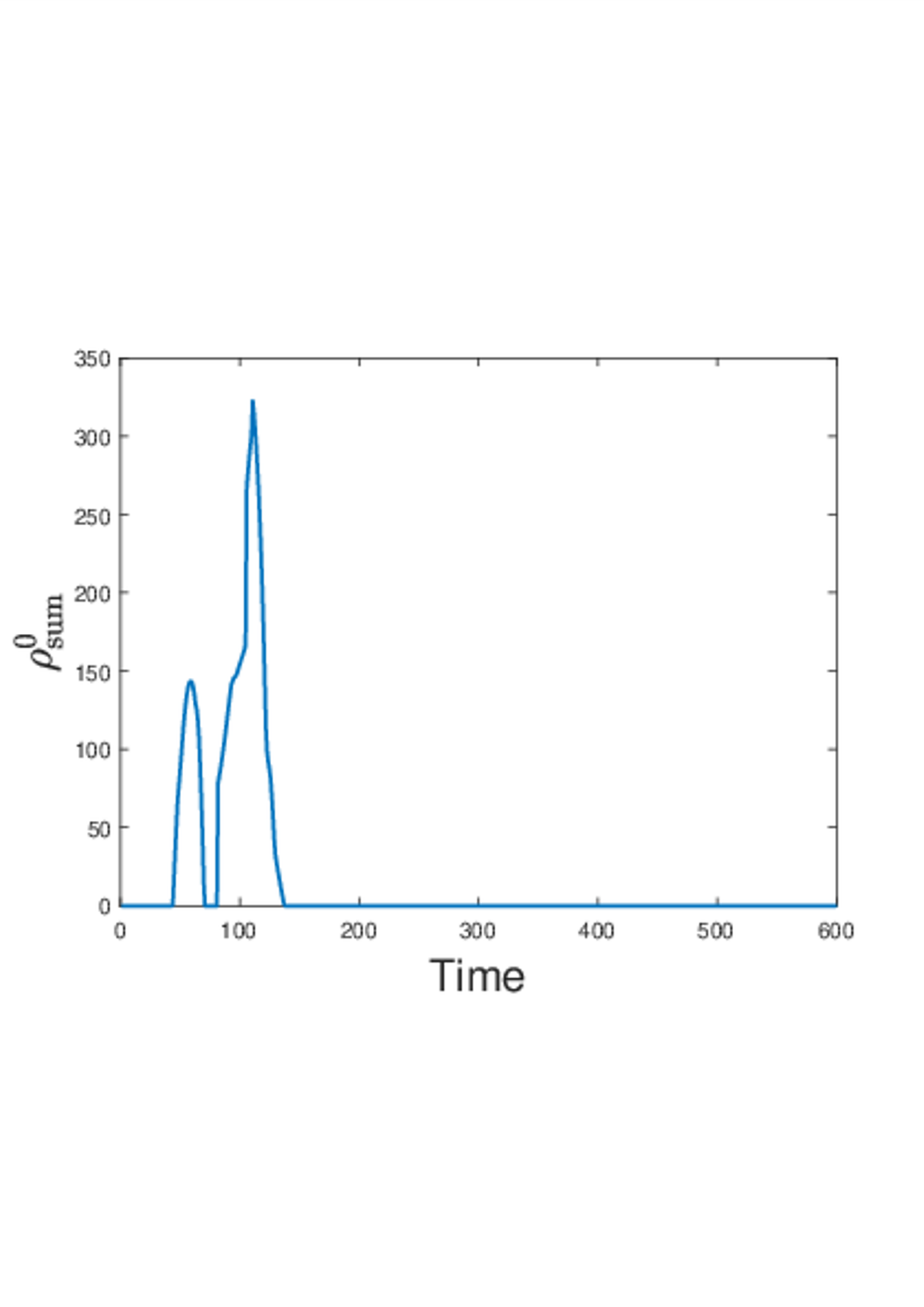}
     \caption{$\rho_{\mathrm{sum}}^0(\boldsymbol{x})$ along the trajectory}
     \label{fig:rho0}
   \end{subfigure}
   \hfill
   \begin{subfigure}[t]{0.23\textwidth}
     \centering
     \includegraphics[width=\textwidth]{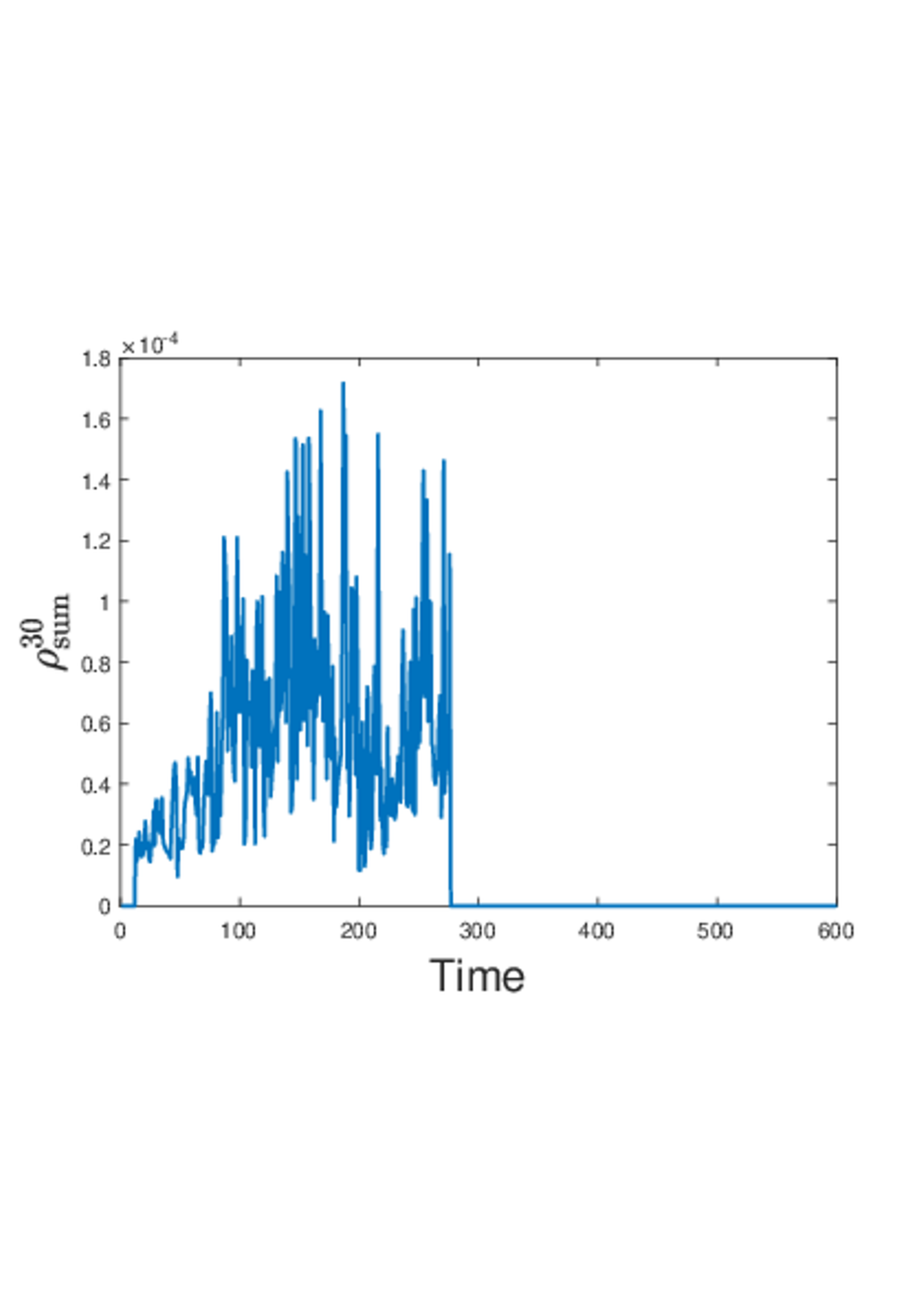}
     \caption{$\rho_{\mathrm{sum}}^{30}(\boldsymbol{x})$} along the trajectory
     \label{fig:rho30}
   \end{subfigure}
   \caption{Evolution of the relaxation parameters $\rho_{\mathrm{sum}}^0(\boldsymbol{x})$ and $\rho_{\mathrm{sum}}^{30}(\boldsymbol{x})$ evaluated at the state trajectory, across algorithm iterations.} 
\end{figure}

The resulting swapping trajectories are shown in Figure \ref{fig:truncontrol}. { Define
\begin{align}
\rho_{\mathrm{sum}}^k&=\sum_{i=1}^N\sum_{e\in\mathcal{C}_i}\left((\rho_{ie}^k)^2+M_i\rho_{ie}^k\right).\label{eq:quan2}
\end{align}}
\noindent The evolution of the relaxation parameters $\rho_{\mathrm{sum}}^0(\boldsymbol{x})$ and $\rho_{\mathrm{sum}}^{30}(\boldsymbol{x})$ { at each time step along the trajectory} is shown in Figures \ref{fig:rho0} and \ref{fig:rho30}. { It can be seen that $\rho_{\mathrm{sum}}^{30}$ is close to zero at every time step, even $\rho_{\mathrm{sum}}^0$ is relatively large at some time steps. This empirically demonstrates the safety guarantees performance of the proposed distributed algorithm. From our experience, $\eta$ could be much smaller for a practical implementation.}

\subsection{Distributed Safety Verification}
{The proposed safety verification procedure is illustrated on a four-robot system within the working space 
\begin{align*}
\mathcal{X} &= \{\boldsymbol{p} \in \mathbb{R}^2:||\boldsymbol{p}||\leq p^{\mathrm{\max}}=4\}\\
&\times \{\boldsymbol{v} \in \mathbb{R}^2:||\boldsymbol{v}||\leq v^{\mathrm{\max}}=3\}. 
\end{align*}
Each robot employs Algorithm \ref{al:truncated} to safely move towards the origin.

We first examine the effect of the number of scenarios $R$ on the verification result. Let $\eta=30$ for the four agents and consider $R=2000$. Each agent maintains a confidence level ${\beta_i}=0.025$ for $i \in \{1,\ldots,4\}$. By solving the verification program \eqref{eq:dscenvarification} and applying Theorem \ref{th:probability}, we establish that with confidence at least $1-\beta=0.9$, the violation probability satisfies $\underline{\epsilon}=0.02\leq\mathbb{P}\left\{\boldsymbol{x}\in \mathcal{H}:\boldsymbol{z}^*=0\notin\mathcal{Z}_{\boldsymbol{x}} \right\}\leq \overline{\epsilon}=0.19$. In practice, an even smaller $\beta_i$ can be chosen; this would only have a mild effect on the interval $[\underline{\epsilon},\overline{\epsilon}]$ due to the way this depends on the confidence.

To validate the obtained probabilistic result, we run $100$ independent experiments, each with $R_{v}=50$ scenarios $\boldsymbol{x}^{(1),j},\ldots,\boldsymbol{x}^{(R_{v}),j}$, for $j\in\{1,\ldots,100\}$. In each experiment $j \in \{1,\ldots,100\}$, we monitor the frequency of violation $f_{v}^j$ by 
\begin{equation}\label{eq:violation_frequency}
f_{v}^j=\sum_{i=1}^{R_{v}}I_{\mathcal{Z}^c_{\boldsymbol{x}^{(i),j}}}(\boldsymbol{z}^*).
\end{equation}
In the above equation, $I_{\mathcal{Z}^c_{\boldsymbol{x}^{(i),j}}}(\boldsymbol{z}^*)$ is the indicator function that equals one if $\boldsymbol{z}^*$ belongs to the set complement of $\mathcal{Z}_{\boldsymbol{x}^{(i),j}}$, i.e., if $\boldsymbol{z}^*\notin\mathcal{Z}_{\boldsymbol{x}^{(i),j}}$, and zero otherwise. For each experiment $j\in\{1,\ldots,100\}$, the violation
probability $\mathbb{P}\{\boldsymbol{x}\in{ \mathcal{H}}:\boldsymbol{z}^*\notin\mathcal{Z}_{\boldsymbol{x}} \}$ is empirically calculated as
\begin{equation}
    \hat{\mathbb{P}}^j\{\boldsymbol{x}\in{ \mathcal{H}}:\boldsymbol{z}^*\notin\mathcal{Z}_{\boldsymbol{x}} \}=\frac{f_{v}^j}{R_{v}}.
\end{equation}

Figure \ref{fig:bargraph} illustrates the bar graph of $f_{v}^j$, $j\in\{1,\ldots,100\}$, while with dashed lines we highlight the theoretical bounds $[\underline{\epsilon},\overline{\epsilon}]$. It can be observed that most of the empirical mass of the violation probability lies between $[\underline{\epsilon},\overline{\epsilon}]$. This in turn implies that our bound for this case study offers a tight estimate of the violation probability.
\begin{figure}[ht]
    \centering
    \includegraphics[width=0.8\linewidth]{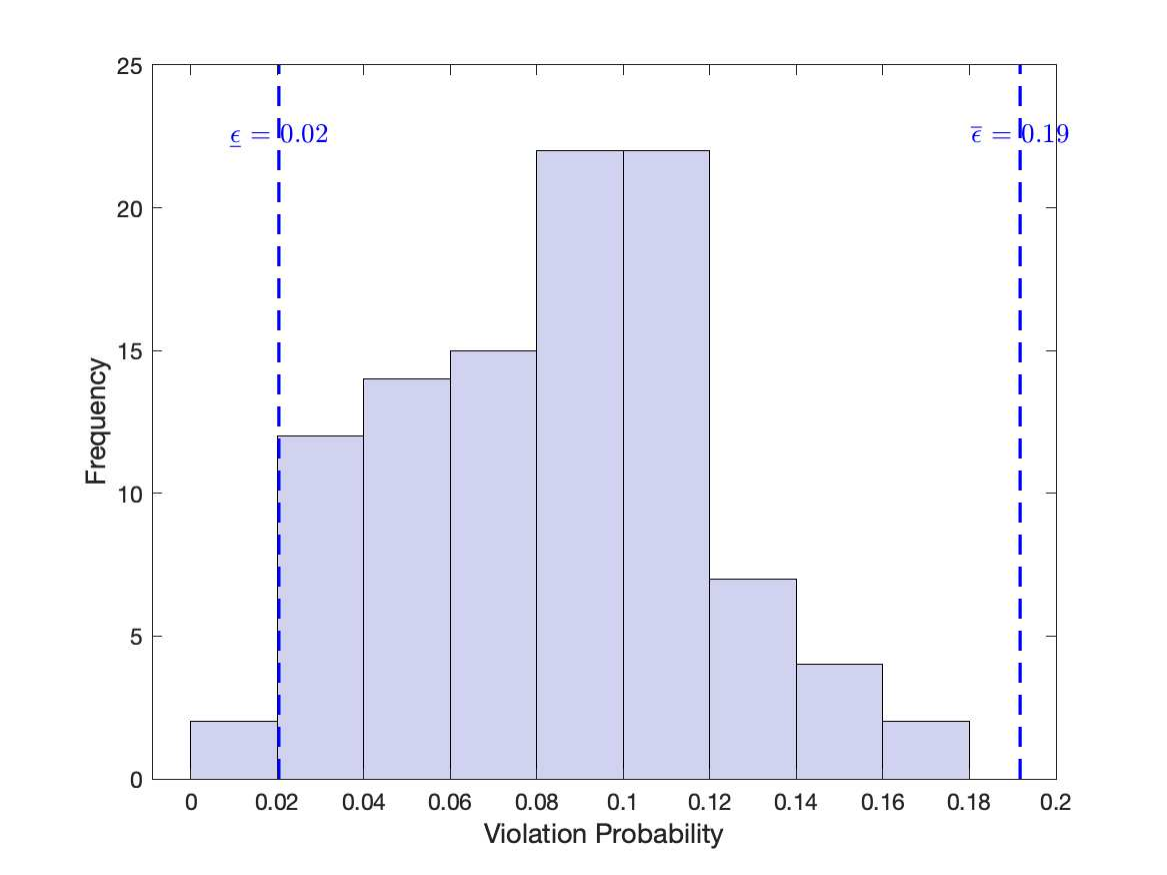}
    \caption{{Bar graph for the violation probability, and (with dashed lines) the theoretical bounds $[\underline{\epsilon},\overline{\epsilon}]$.}}
    \label{fig:bargraph}
\end{figure}

The empirical cumulative distribution function (CDF) of $\mathbb{P}\left\{\boldsymbol{x}\in{ \mathcal{H}}:\boldsymbol{z}^*\notin\mathcal{Z}_{\boldsymbol{x}} \right\}$ can be constructed using $\hat{\mathbb{P}}^j\{\boldsymbol{x}\in{ \mathcal{H}}:\boldsymbol{z}^*\notin\mathcal{Z}_{\boldsymbol{x}} \}$, $j\in\{1,\ldots,100\}$.
These results are shown in Figure \ref{fig:distribution};
it can be observed that the empirical probability implies that $\mathbb{P}\{\mathbb{P}\left\{\boldsymbol{x}\in{ \mathcal{H}}:\boldsymbol{z}^*\notin\mathcal{Z}_{\boldsymbol{x}} \right\} \in [\underline{\epsilon},\overline{\epsilon}]\} \approx 0.9134 \ge 1-\beta=0.9$, thus demonstrating numerically the confidence of the theoretical result of Theorem \ref{th:probability}. 
\begin{figure}[ht]
   \centering
   \begin{subfigure}[b]{0.35\textwidth}
     \centering
     \includegraphics[width=1.2\textwidth]{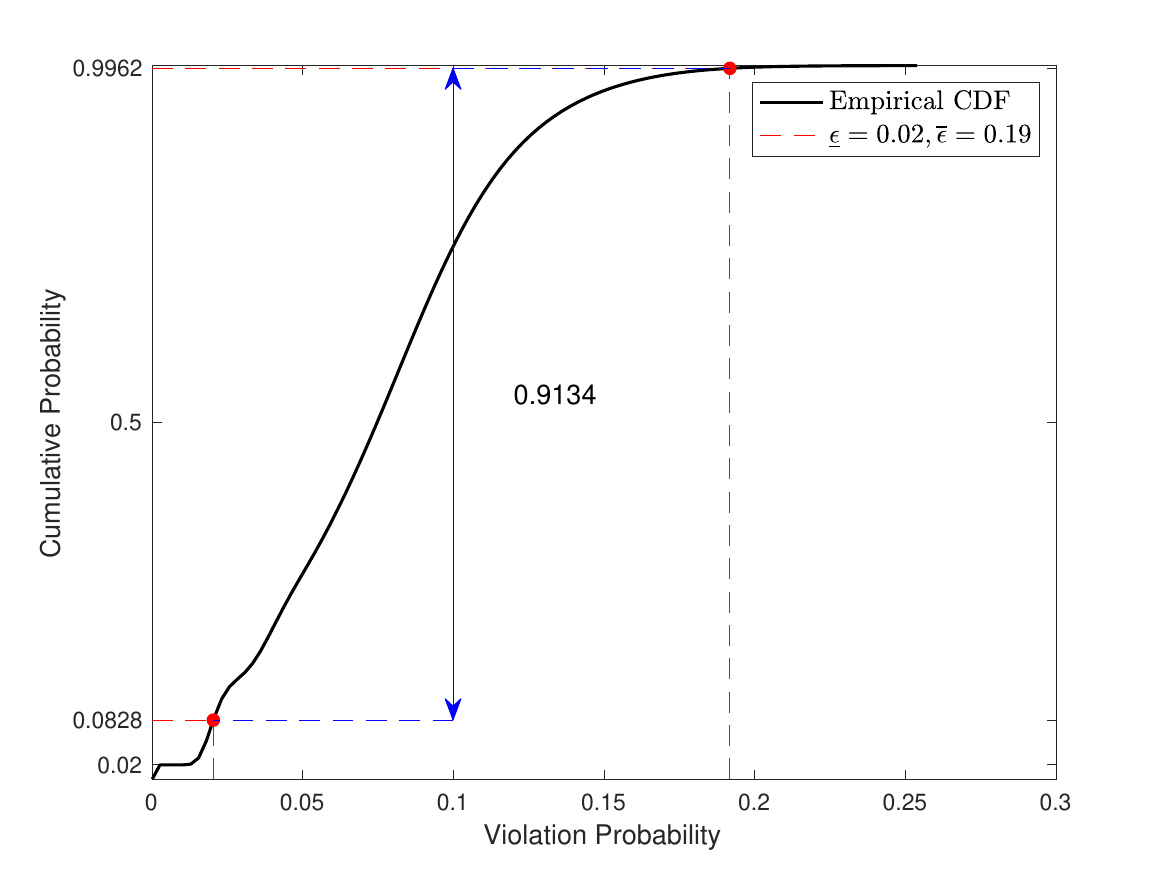}
    % \caption{CDF for safety violation}
     \label{fig:4cdf}
   \end{subfigure}
   \caption{{Empirical Cumulative distribution function (CDF) for $\mathbb{P}\left\{\boldsymbol{x}\in{ \mathcal{H}}:\boldsymbol{z}^*\notin\mathcal{Z}_{\boldsymbol{x}} \right\}$, theoretic bounds $[\underline{\epsilon},\overline{\epsilon}]$ and the corresponding empirical probability. The horizontal axis represents the violation probability while the vertical axis represents the associated empirical cumulative probability.}}
     \label{fig:distribution}
\end{figure}

% In simulation it is also observed that with larger $R$ such as $R=5000$, the bounds remain almost the same as these for $R=2000$. Finding a small but efficient $R$ will be explored in future research. 
}
\section{Conclusion}
\label{sec:conc}
In this paper we presented distributed safe control design and safety verification algorithms for multi-agent systems. The proposed control algorithms introduce auxiliary and relaxation variables to allow feasibility across iterations. We guaranteed convergence to an optimal solution and established a sublinear convergence rate under certain conditions. We also addressed the problem of distributed safety verification for given control inputs. A scenario-based verification program was formulated and can be solved locally by each agent. The scenarios are sampled independently by a sequential algorithm. The distributed scenario program characterizes the probability of being unsafe, with both lower and upper bounds being determined. Simulation on a multi-robot swapping position problem demonstrated the efficacy of our result. Current work concentrates in accounting for communication delays and model uncertainty in real systems.

\appendix
\section{Appendix}
\begin{pf}[Proof of Lemma \ref{lem:newproperty}]
% Combining the optimisation problems \eqref{eq:dcbf} in Step 3 of Algorithm \ref{al:dcbf} for all agents, we have
% \begin{equation}\label{eq:globalrelaxedcbf}
%   \begin{split}
%     \min_{\boldsymbol{u},\boldsymbol{\rho}}&~J(\boldsymbol{u},\boldsymbol{\rho})\\
%     \mathrm{subject~to}~&u_i\in\mathcal{U}_i,\forall i=1,\ldots,N,\boldsymbol{\rho}\ge 0\\
%     &\sum_{k\in\mathcal{V}_e} h_{ke}(u_k)\le\sum_{k\in\mathcal{V}_e}\rho_{ke},\forall e=1,\ldots,E,
%   \end{split}
% \end{equation}
% where for each $e=1,\ldots,E$, $\mu_e$ is the dual variable associated with the last constraint in \eqref{eq:globalrelaxedcbf}.
The dual function of the relaxed problem \eqref{eq:relaxedqp} is given by
\begin{align}\label{eq:dual}
    q(\boldsymbol{\mu})&:=\inf_{\{u_i\in\mathcal{U}_i\}_{i=1}^N,\boldsymbol{\rho}\ge 0}\sum_{i=1}^N\left\{J_i(u_i)+\sum_{e\in\mathcal{C}_i}(\rho^2_{ie}+M_{i}\rho_{ie})\right\}\nonumber\\
    &+\sum_{e=1}^E\mu_e\left\{\sum_{i\in\mathcal{V}_e} h_{ie}(u_i)-\sum_{i\in\mathcal{V}_e}\rho_{ie}\right\}\nonumber\\
    &=\inf_{\{u_i\in\mathcal{U}_i\}_{i=1}^N,\boldsymbol{\rho}\ge 0}\sum_{i=1}^N\left\{J_i(u_i)+\sum_{e\in\mathcal{C}_i}\mu_eh_{ie}(u_i)\right\}\nonumber\\
    &+\sum_{e=1}^E\sum_{i\in\mathcal{V}_e}\left\{\rho_{ie}^2+(M_{i}-\mu_e)\rho_{ie}\right\}.
    \nonumber\\
    &=q_{\mathrm{nom}}(\boldsymbol{\mu})+\inf_{\boldsymbol{\rho}\ge 0}\sum_{e=1}^E\sum_{i\in\mathcal{V}_e}\left\{\rho_{ie}^2+(M_{i}-\mu_e)\rho_{ie}\right\},
\end{align}
where 
\begin{equation}\label{eq:A.2}
    q_{\mathrm{nom}}(\boldsymbol{\mu}):=\inf_{\{u_i\in\mathcal{U}_i\}_{i=1}^N}\sum_{i=1}^N\left\{J_i(u_i)+\sum_{e\in\mathcal{C}_i}\mu_eh_{ie}(u_i)\right\}
\end{equation}
is the dual function of the nominal problem \eqref{eq:centralizedqp}. Let $\boldsymbol{\mu}^*$ be the maximizer for $q(\boldsymbol{\mu})$, and $\tilde{\boldsymbol{\mu}}$ be a maximizer of $q_{\mathrm{nom}}(\boldsymbol{\mu})$.

{Now consider the value of the second term in $q(\boldsymbol{\mu})$. If $\mu_e\le M_i$ for some $e\in\{1,\ldots,E\},i\in\mathcal{V}_e$, then
\begin{equation}\label{eq:case1}
    \inf_{\rho_{ie}\ge 0}\rho_{ie}^2+(M_i-\mu_e)\rho_{ie}=0,\rho_{ie}^*=0.
\end{equation}
Otherwise if $\mu_e>M_i$, then
\begin{equation}\label{eq:case2}
    \inf_{\rho_{ie}\ge 0}\rho_{ie}^2+(\mu_e-M_i)\rho_{ie}=-\frac{(M_i-\mu_e)^2}{4},\rho_{ie}^*=\frac{\mu_e-M_i}{2}.
\end{equation}
We first show that if \eqref{eq:lem2eq1} holds, then 
\begin{equation}\label{eq:done}
    M_{i}\ge{\mu}_{e}^*,\forall i\in\mathcal{V}_e, \forall e\in\{1,\ldots,E\},
\end{equation}
where $\mu_e^*$ is the $e$-th element of $\boldsymbol{\mu}^*$. Suppose for the sake of contradiction that there exists $e\in\{1,\ldots,E\},i\in\mathcal{V}_e$ such that $\mu_e^*> M_i$. Then, from \eqref{eq:dual}, \eqref{eq:case1} and \eqref{eq:case2} we have
\begin{equation*}
    q({\boldsymbol{\mu}}^*)<q_{\mathrm{nom}}({\boldsymbol{\mu}}^*)\le q_{\mathrm{nom}}(\tilde{\boldsymbol{\mu}}^*).
\end{equation*}
The second inequality comes from a fact that $\tilde{\boldsymbol{\mu}}^*$ is a maximizer of $q_{\mathrm{nom}}(\boldsymbol{\mu})$. However, from \eqref{eq:dual}, \eqref{eq:case1}, and {\eqref{eq:lem2eq1}} we have
\begin{equation*}
    q(\tilde{\boldsymbol{\mu}}^*)=q_{\mathrm{nom}}(\tilde{\boldsymbol{\mu}}^*).
\end{equation*}
We conclude that $q(\tilde{\boldsymbol{\mu}}^*)>q(\boldsymbol{\mu}^*)$, thus reach a contradiction as $\boldsymbol{\mu}^*$ maximizes $q(\boldsymbol{\mu})$. By \eqref{eq:done} and \eqref{eq:case1} we have $\boldsymbol{\rho}^*=0$, any $\tilde{\boldsymbol{\mu}}^*$ that maximizes $q_\mathrm{nom}(\boldsymbol{\mu})$ also maximizes $q({\boldsymbol{\mu}})$.

As a direct result of \eqref{eq:case1} and \eqref{eq:case2}, the second part of $q(\boldsymbol{\mu})$, $\inf_{\boldsymbol{\rho}\ge 0}\sum_{e=1}^E\sum_{i\in\mathcal{V}_e}\left\{\rho_{ie}^2+(M_{i}-\mu_e)\rho_{ie}\right\}$, is concave and smooth. This is different from \cite[Lemma III.2]{notarnicola2019constraint} where the dual function goes to $-\infty$ when $\mu_e>M_i.$ Introducing a quadratic term for the relaxation variables enhances convexity of the primal function, hence smoothness of the dual function.} 

\end{pf}
\begin{pf}[Proof of Theorem \ref{th:speed}]
We begin with (a). 
% The proof follows a similar idea as \cite[Theorem II.6]{notarstefano2019distributed}, we only scratch the primal-dual exploration here. 
Under Assumption \ref{ass:slater}, strongly duality holds for the primal problem \eqref{eq:centralizedqp} and the dual problem \eqref{eq:dual}. { With a slight abuse of notation, we define
\begin{align}\label{eq:newdual}
    q_i(\boldsymbol{\mu}_i):=&\inf_{\{u_i\in\mathcal{U}_i\},\boldsymbol{\rho}\ge 0}\left\{J_i(u_i)+\sum_{e\in\mathcal{C}_i}(\rho^2_{ie}+M_{i}\rho_{ie})\right.\nonumber\\
    &+\left.\sum_{e\in\mathcal{C}_i}\mu_{ie}(h_{ie}(u_i)-\rho_{ie})\right\}.
\end{align}
By Assumption \ref{ass:connectivity}, we have $\mathcal{G}_e$ is undirected and connected for every $e\in\{1,\ldots,E\}$. Therefore, suppose  $\mu_{ie}={\mu_{le}}$, $\forall e\in\{1,\ldots,E\},i\in\mathcal{V}_e$, $l\in\mathcal{N}_i\cap\mathcal{V}_e$, then we can deduce that
\begin{equation}\label{eq:newcon}
    \mu_{ie}={\mu_{le},\forall i,l\in\mathcal{V}_e.}
\end{equation}
Recalling that $i\in\{1,\ldots,N\}$ is the numbering of agent, $e\in\mathcal{C}_i$ is the {numbering} of CBF constraint that involves agent $i$, $\mathcal{N}_i$ is the set of neighbouring agents for agent $i$, and $\mathcal{V}_e$ is the set of agents in {sub-network} $\mathcal{G}_e$. $\mathcal{N}_i\cap\mathcal{V}_e\ne\emptyset$ due to Assumption \ref{ass:connectivity}. The new variable $\mu_{ie}$ and $\mu_{le}$ can be regarded as local copies of $\mu_e$ by agent $i$ and agent $l$, which are associated with the $e$-th CBF constraint. Using the decomposed dual function \eqref{eq:newdual} and the new constraint \eqref{eq:newcon}, we come up with an equivalent decomposed dual problem
\begin{equation}\label{eq:dualdecomposition}
  \begin{split}
    \max_{\boldsymbol{\mu}_i\ge 0}~&\sum_{i=1}^Nq_i(\boldsymbol{\mu_i})\\
    \mathrm{subject~to}~&\mu_{ie}=\mu_{le},\forall i\in\{1,\ldots,N\},e\in\mathcal{C}_i,l\in\mathcal{N}_i\cap\mathcal{V}_e,
  \end{split}
\end{equation}
If $\mathcal{V}_e=\{1,\ldots,N\},\forall e\in\{1,\ldots,E\}$, \eqref{eq:dualdecomposition} is a generic dual decomposition problem \cite[Section 3.1.3]{notarstefano2019distributed}. 

Consider the dual function of \eqref{eq:dualdecomposition}
\begin{equation}\label{eq:dualdualdual}
    d(\boldsymbol{\lambda}):=\sum_{i=1}^N\sup_{\boldsymbol{\mu}_i\ge 0}\left(q_i(\boldsymbol{\mu}_i)+\sum_{e\in\mathcal{C}_i}\sum_{l\in\mathcal{N}_i\cap\mathcal{V}_e}\lambda_{il}^\top(\mu_{ie}-\mu_{le})\right),
\end{equation}
where $\lambda_{il}$ is a free dual variable for the constraint $\mu_{ie}=\mu_{le}$ in \eqref{eq:dualdecomposition}. Recalling that the network $\mathcal{G}$ is undirected, for each $(i,l)\in\mathcal{E}$ we also have $(l,i)\in\mathcal{E}$. This indicates that in \eqref{eq:dualdualdual}, we have both $\lambda_{il}^\top(\mu_{ie}-\mu_{le})$ and $\lambda_{li}^\top(\mu_{le}-\mu_{ie})$ for every given $i\in\{1,\ldots,N\},e\in\mathcal{C}_i,l\in\mathcal{N}_i\cap\mathcal{V}_e$. By gathering the terms involving $\boldsymbol{\mu}_i$ together, such as $\lambda_{il}^\top \mu_{ie}$ and $-\lambda_{li}^\top \mu_{ie}$, and doing some algebraic calculations, we obtain
\begin{equation}\label{eq:dualdualfunc}
d(\boldsymbol{\lambda})=\sum_{i=1}^N\sup_{\boldsymbol{\mu}_i\ge 0}\left(q_i(\boldsymbol{\mu}_i)+\sum_{e\in\mathcal{C}_i}\mu_{ie}^\top\sum_{l\in\mathcal{N}_i\cap\mathcal{V}_e}(\lambda_{il}-\lambda_{li})\right)
\end{equation}
As \eqref{eq:dualdualdual} is traversing every $i\in\{1,\ldots,N\}$, $\mu_{le}$ in \eqref{eq:dualdualdual} is also contained in \eqref{eq:dualdualfunc}, for $l\in\mathcal{V}_e$. A procedure similar to \eqref{eq:dualdualdual} and \eqref{eq:dualdualfunc} has been proposed in \cite[Section III.B]{notarnicola2019constraint} but only for one network $\mathcal{G}$. Our formulation generalizes these results to constraints defined on multiple sub-networks $\mathcal{G}_e$, for $e\in\{1,\ldots,E\}.$
}

The dual problem of \eqref{eq:dualdecomposition} is then given by
\begin{equation}\label{eq:dualdualdecomposition}
  \begin{split}
    d^*=\min_{\boldsymbol{\lambda}}d(\boldsymbol{\lambda}).
  \end{split}
\end{equation}
Strong duality holds between problem \eqref{eq:dualdecomposition} and \eqref{eq:dualdualdecomposition} since \eqref{eq:dualdecomposition} is an linear equality constrained concave problem. {Therefore, solving problem \eqref{eq:dualdualdecomposition} leads to the optimal solution of problem \eqref{eq:dualdecomposition}. Solving problem \eqref{eq:dualdualdecomposition} has advantages in terms of distributed computation. This can be seen by applying the gradient descent method to solve \eqref{eq:dualdualdecomposition}. From \eqref{eq:dualdualdual}, for every $i\in\{1,\ldots,N\}$, $e\in\mathcal{C}_i$, and $l\in\mathcal{N}_i\cap\mathcal{V}_e$, the gradient $\nabla d(\lambda_{il})$ is given by
\begin{equation}\label{eq:gd}
    \nabla d(\lambda_{il})=\mu_{ie}-\mu_{le}.
\end{equation}
} At iteration $k$, each agent $i$ performs two steps: \begin{itemize}
  \item (i) {for every $e\in\mathcal{C}_i$, $l\in\mathcal{N}_i\cap\mathcal{V}_e$, calculate the gradient $\nabla d(\lambda_{il}^k)$: receive $\lambda_{li}^k$, $l\in\mathcal{N}_i\cap\mathcal{V}_e$, and compute $\mu_{ie}$ by solving 
  \begin{equation}\label{eq:relaxeddual}
    \max_{\boldsymbol{\mu}_i\ge 0}\left(q_i(\boldsymbol{\mu}_i)+\sum_{e\in\mathcal{C}_i}\mu_{ie}^\top\sum_{l\in\mathcal{N}_i\cap\mathcal{V}_e}(\lambda_{il}^k-\lambda_{li}^k)\right).
  \end{equation}}
  \item (ii) use gradient descent: for every $e\in\mathcal{C}_e,l\in\mathcal{N}_i\cap\mathcal{V}_e$, receive $\mu_{le}^k$ and update $\lambda_{il}$ by \eqref{eq:gd}:
  \begin{equation}\label{eq:gradientdescent}
    \lambda_{il}^{k+1}=\lambda_{il}^k-\gamma^{k}(\mu_{ie}^k-\mu_{le}^{k}).
  \end{equation}
\end{itemize}

{\eqref{eq:gradientdescent} is Step 5 of Algorithm \ref{al:dcbf}. We then show that solving \eqref{eq:relaxeddual} is equivalent to solving \eqref{eq:dcbf} at Step 3.
For every $i\in\{1,\ldots,N\}$, dualizing the CBF constraints in \eqref{eq:dcbf} by $\boldsymbol{\mu}_i\ge 0$ yields a dual problem
\begin{align}
    &\max_{\boldsymbol{\mu}_i\ge 0}\inf_{\{u_i\in\mathcal{U}_i\},\boldsymbol{\rho}\ge 0}\left\{J_i(u_i)+\sum_{e\in\mathcal{C}_i}(\rho^2_{ie}+M_{i}\rho_{ie})\right.\nonumber\\
    &+\left.\sum_{e\in\mathcal{C}_i}\mu_{ie}(h_{ie}(u_i)-\rho_{ie})\right\}+\sum_{e\in\mathcal{C}_i}\mu_{ie}^\top \sum_{l\in\mathcal{N}_i\cap\mathcal{V}_e}(\lambda_{il}^k-\lambda_{li}^k)\nonumber\\
    &\stackrel{\eqref{eq:newdual}}{=}\max_{\boldsymbol{\mu}_i\ge 0}\left(q_i(\boldsymbol{\mu}_i)+\sum_{e\in\mathcal{C}_i}\mu_{ie}^\top\sum_{l\in\mathcal{N}_i\cap\mathcal{V}_e}(\lambda_{il}^k-\lambda_{li}^k)\right),\label{eq:dualoptimization}
\end{align}
which is \eqref{eq:relaxeddual}. Therefore, Steps 2-5 in Algorithm \ref{al:dcbf} involve performing gradient descent to solve problem \eqref{eq:dualdualdecomposition} in a distributed manner.
}

Diminishing step-size is used here as \cite{notarnicola2019constraint}. Specifically, \eqref{eq:relaxeddual} is the dual problem of \eqref{eq:dcbf}. Strong duality holds for large enough $\boldsymbol{\rho}$ as the relaxed CBF constraints hold strictly. Updating \eqref{eq:gradientdescent} is the same as \eqref{eq:trunsubupdate} for every agent across iterations. Given that $d(\boldsymbol{\lambda})$ is convex, gradient descent guarantees that $d(\boldsymbol{\lambda}^k)$ convergence to the optimal value $d^*=J^*$ since strong duality holds between \eqref{eq:relaxedqp} and \eqref{eq:dualdecomposition}, as well as \eqref{eq:dualdecomposition} and \eqref{eq:dualdualdecomposition}. Moreover, the relaxed problem \eqref{eq:relaxedqp} is strongly (hence also strictly) convex, which indicates uniqueness of the optimal solution $(\boldsymbol{u}_{\mathrm{rel}}^*,\boldsymbol{\rho}^*)$ . Using Lemma \ref{lem:newproperty}, we obtain $\boldsymbol{u}_{\mathrm{rel}}^*=\boldsymbol{u}_{\mathrm{nom}}^*$, which is the optimal solution of \eqref{eq:centralizedqp}.

We then prove $(b)$. First we prove that $q(\boldsymbol{\mu})$ in \eqref{eq:dual} is a concave quadratic function. When every $\mathcal{U}_i=\mathbb{R}^{m_i}$ and the CBF constraints are linearly independent, the relaxed CBF-QP \eqref{eq:relaxedqp} is a linearly constrained strongly convex quadratic problem. Following the example \cite[Section 5.2.4, Eq. 5.28]{boyd2004convex}\footnote{The example demonstrates that the dual function of a convex quadratically constrained quadratic programming problem is a concave quadratic function. Our problem is as a special case where the quadratic terms are zero in the constraints. }, {$q_{\mathrm{nom}}(\boldsymbol{\mu})$ in \eqref{eq:A.2} is a strongly concave quadratic function. Together with \eqref{eq:dual}, \eqref{eq:case1} and \eqref{eq:case2}, we conclude that $q(\boldsymbol{\mu})$ is a strongly concave and smooth function.}
% with the quadratic part as $-\boldsymbol{\mu_i}^\top G_i G_i^\top\boldsymbol{\mu_i}$, where
% \begin{equation}\label{eq:quadraticmatrix}
% {G_i[i,e]} = \left\{ \begin{array}{l}
% {{\mathcal L}_{{g_i}}}{b_e},~\text{if $i\le N,i\in\mathcal{V}_e$}\\
% - 1~~~~~\text{if $i>N,i\in\mathcal{V}_e$}\\
% 0~~~~~~~\text{if $i>N,i\notin\mathcal{V}_e$}
% \end{array} \right.
% \end{equation}
% For $i\le N,i\in\mathcal{V}_e$, $G_{ie}$ corresponds to the coefficient for $u_{ie}$ in the linear constraints of \eqref{eq:relaxedqp}. For $e>E,e\in\mathcal{C}_i$, $G_{ie}$ corresponds to the coefficient, which is $-1$ for the relaxation variable $\rho_{i(e-E)}$.
{From duality between strong concavity (convexity) and smoothness \cite[Theorem 6]{kakade2009duality}, $d(\boldsymbol{\lambda})$ is a smooth and necessarily convex function.} Using constant step size
\begin{equation}\label{eq:stepsize}
  0 < \gamma < \frac{1}{2L},
\end{equation}
where $L$ is Lipschitz constant of $\nabla d(\boldsymbol{\lambda}),$ in a gradient descent method to minimize a smooth and convex function $d(\boldsymbol{\lambda})$, the generated iterates converge sublinearly as
\begin{align}\label{eq:sublinearrate}
  d(\boldsymbol{\lambda}^{k})-J^*&\le \frac{2(d(\boldsymbol{\lambda}^0)-J^*)||\boldsymbol{\lambda^0}-\boldsymbol{\lambda^*}||_2^2}{2||\boldsymbol{\lambda}^0-\boldsymbol{\lambda}^*||_2^2+k\gamma(2-L\gamma)(d(\boldsymbol{\lambda}^0)-J^*)}\nonumber\\
  \le &\frac{2(d(\boldsymbol{\lambda}^0)-J^*)||\boldsymbol{\lambda^0}-\boldsymbol{\lambda^*}||_2^2}{k\gamma(d(\boldsymbol{\lambda}^0)-J^*)}\le \frac{2||\boldsymbol{\lambda}^0-\boldsymbol{\lambda}^*||_2^2}{k\gamma}.
\end{align}
The first inequality is proved by \cite[Theorem 2.1.14]{nesterov2003introductory}, the second one comes from eliminating the term $||\boldsymbol{\lambda}^0-\boldsymbol{\lambda}^*||_2^2$ from the denominator, and considering $2-L\gamma\ge 1$ from \eqref{eq:stepsize}.

{Recalling the expression of $d(\boldsymbol{\lambda})$ from \eqref{eq:newdual}, and the duality result from \eqref{eq:dualoptimization}, we have
\begin{align}\label{eq:primaldual}
&d(\boldsymbol{\lambda}^k)=\nonumber\\
&\sum_{i=1}^N\inf_{u_i,\boldsymbol{\rho}_i\ge 0}\left(\sup_{\boldsymbol{\mu}_i\ge 0}\left(J_i(u_i)+\sum_{e\in\mathcal{C}_i}\left(\rho_{ie}^2+M_i\rho_{ie}\right)\right)\right.\nonumber\\
&\left.+\sum_{e\in\mathcal{C}_i}\mu_{ie}(h_{ie}(u_i)-\rho_{ie})+\sum_{e\in\mathcal{C}_i}\mu_{ie}^\top\sum_{l\in\mathcal{N}_i\cap\mathcal{V}_e}(\lambda_{il}^k-\lambda_{li}^k)\right)\nonumber\\
&=\sum_{i=1}^N\left(J_i(u_i^k+\sum_{e\in\mathcal{C}_i}((\rho_{ie}^k)^2+M_i\rho_{ie}^k)\right)\nonumber\\
&=\sum_{i=1}^N||u_i^{k}-u^{\mathrm{des}}||^2+\rho_{\mathrm{sum}}^{k}=H(\boldsymbol{u}^k,\boldsymbol{\rho}^k).
\end{align}
Hence, by \eqref{eq:sublinearrate} and \eqref{eq:primaldual}, we conclude that $H(\boldsymbol{u}^k,\boldsymbol{\rho}^k)-J^*<\frac{2||\boldsymbol{\lambda}^0-\boldsymbol{\lambda}^*||_2^2}{\gamma k}$.

% Under certain regularity condition, \cite[Algorithm RSDD]{notarnicola2019constraint} only guarantees local sublinear convergence \cite{camisa2021distributed} since $\nabla d(\boldsymbol{\lambda})$ is only guaranteed to be bounded \cite[Proposition 5.2]{camisa2021distributed}. Lipschitz continuity of $\nabla d(\boldsymbol{\lambda})$ is important to establish a global sublinear convergence rate.
}
\end{pf}

\begin{pf}[Proof of Theorem \ref{th:probability}]
We have that 
\begin{align}\label{eq:scinequality1}
   \mathbb{P}^{{ R}}&\left\{\frac{\sum_{i=1}^N\underline\epsilon_i(s_i^*)}{N} \le\mathbb{P}\left\{\boldsymbol{x}\in{ \mathcal{H}}:\boldsymbol{z}^*\notin\mathcal{Z}_{\boldsymbol{x}} \right\} \le \sum_{i=1}^N\bar\epsilon_i(s_i^*)
  \right\}\nonumber\\
  &=\mathbb{P}^{{ R}}\left\{\frac{1}{N}\sum_{i=1}^N\underline\epsilon_i(s_i^*)\le\mathbb{P}\left\{\vphantom{\bigcap_{i=1}^N}\boldsymbol{x}\in{ \mathcal{H}}: \right.\right.\nonumber\\
  &\hspace{0.1cm}\exists i\in\{1,\ldots,N\},\left.\left.\vphantom{\bigcap_{i=1}^N}\boldsymbol{z}^*\notin{\mathcal{Z}_{\boldsymbol{x}}^i} \right\}\le\sum_{i=1}^N\bar\epsilon_i(s_i^*)\right\}\nonumber\\
  &=\mathbb{P}^{{ R}}\left\{\frac{1}{N}\sum_{i=1}^N\underline\epsilon_i(s_i^*)\le\mathbb{P}\left\{\boldsymbol{x}\in{ \mathcal{H}}:\bigcup_{i=1}^N\left\{\boldsymbol{z}^*\notin{\mathcal{Z}_{\boldsymbol{x}}^i} \right\}\right\}\right.\nonumber\\
  &\bigcap\left.\mathbb{P}\left\{\bigcup_{i=1}^N\left\{\boldsymbol{x}\in{ \mathcal{H}}:\boldsymbol{z}^*\notin{\mathcal{Z}_{\boldsymbol{x}}^i} \right\}\right\}\le\sum_{i=1}^N\bar\epsilon_i(s_i^*)\right.\}
  \end{align}
  {The second equation comes from the fact that $\boldsymbol{z}^*\in\mathcal{Z}_{\boldsymbol{x}}$ is equivalent to $\boldsymbol{z}^*\in{\mathcal{Z}_{\boldsymbol{x}}^i}~\forall i\{1,\ldots,N\}.$ The second equation changes $\exists i\in\{1,\ldots,N\},\boldsymbol{z}^*\notin{\mathcal{Z}_{\boldsymbol{x}}^i}$ into $\bigcup_{i=1}^N\{\boldsymbol{z}^*\notin{\mathcal{Z}_{\boldsymbol{x}}^i}\}$. Similar tricks have been used in \cite[Equation 15]{margellos2017distributed} to derive an upper bound for the inner probability. Here we extend the results to both upper and lower bounds, using Theorem \ref{th:scenario}.} We separately deal with the two bounds on the probability. For the upper bound we have
  \begin{align*}
   \mathbb{P}^{{ R}}&\left\{\mathbb{P}\left\{\bigcup_{i=1}^N\left\{\boldsymbol{x}\in{ \mathcal{H}}:\boldsymbol{z}^*\notin{\mathcal{Z}_{\boldsymbol{x}}^i} \right\}\right\}\le\sum_{i=1}^N\bar\epsilon_i(s_i^*)\right\}\\
   \ge &\mathbb{P}^{{ R}}\left\{\sum_{i=1}^N\mathbb{P}\left\{\boldsymbol{x}\in{ \mathcal{H}}:\vphantom{\bigcup}\boldsymbol{z}^*\notin{\mathcal{Z}_{\boldsymbol{x}}^i} \right\}\le\sum_{i=1}^N\bar\epsilon_i(s_i^*)\right\}.
  \end{align*}
  The equality is achieved when for any $i\ne j$, $\boldsymbol{z}^*\notin{\mathcal{Z}_{\boldsymbol{x}}^i}$ and $\boldsymbol{z}^*\notin{\mathcal{Z}_{\boldsymbol{x}}^j}$ are mutually exclusive.
  For the lower bound we have
  \begin{align*}
    \mathbb{P}^{{ R}}&\left\{\frac{1}{N}\sum_{i=1}^N\underline\epsilon_i(s_i^*)\le\mathbb{P}\left\{\boldsymbol{x}\in{ \mathcal{H}}:\bigcup_{i=1}^N\left\{\boldsymbol{z}^*\notin{\mathcal{Z}_{\boldsymbol{x}}^i} \right\}\right\}\right\}\\
    \ge &\mathbb{P}^{{ R}}\left\{N\cdot\frac{1}{N}\sum_{i=1}^N\underline\epsilon_i(s_i^*)\le\sum_{i=1}^N\mathbb{P}\left\{\boldsymbol{x}\in{ \mathcal{H}}:\vphantom{\bigcup}\boldsymbol{z}^*\notin{\mathcal{Z}_{\boldsymbol{x}}^i} \right\}\right\}.
  \end{align*}
  The equality is achieved if for any $i\ne j$, $\boldsymbol{z}^*\ne {\mathcal{Z}_{\boldsymbol{x}}^i}\Leftrightarrow\boldsymbol{z}^*\ne{\mathcal{Z}_{\boldsymbol{x}}^j}$ and $\underline{\epsilon}_i(s_i^*)=\underline{\epsilon}_j(s_j^*)$. The right-hand side of \eqref{eq:scinequality1} can be then lower-bounded by
  \begin{align}
  &\mathbb{P}^{{ R}}\left\{N\cdot\frac{1}{N}\sum_{i=1}^N\underline\epsilon_i(s_i^*)\le\sum_{i=1}^N\mathbb{P}\left\{\boldsymbol{x}\in{ \mathcal{H}}:\vphantom{\bigcup}\boldsymbol{z}^*\notin{\mathcal{Z}_{\boldsymbol{x}}^i} \right\}\right.\nonumber\\&\bigcap
  \left.\sum_{i=1}^N\mathbb{P}\left\{\boldsymbol{x}\in{ \mathcal{H}}:\vphantom{\bigcup}\boldsymbol{z}^*\notin{\mathcal{Z}_{\boldsymbol{x}}^i} \right\}\le\sum_{i=1}^N\bar\epsilon_i(s_i^*)\right\}\nonumber\\
    &\ge \mathbb{P}^{{ R}}\left\{\bigcap_{i=1}^N\left\{\underline\epsilon_i(s_i^*)\le\mathbb{P}\left\{\boldsymbol{x}\in{ \mathcal{H}}:\vphantom{\bigcup}\boldsymbol{z}^*\notin{\mathcal{Z}_{\boldsymbol{x}}^i} \right\}\le\bar\epsilon_i(s_i^*)\right\}\right\}\nonumber\\
    &\ge
    1-\sum_{i=1}^N\mathbb{P}^{{ R}}\left\{\bar\epsilon_i(s_i^*)<\mathbb{P}\left\{\boldsymbol{x}\in{ \mathcal{H}}:\vphantom{\bigcup}\boldsymbol{z}^*\notin{\mathcal{Z}_{\boldsymbol{x}}^i} \right\}\right.\nonumber\\
    &\bigcup\left.\mathbb{P}\left\{\boldsymbol{x}\in{ \mathcal{H}}:\vphantom{\bigcup}\boldsymbol{z}^*\notin{\mathcal{Z}_{\boldsymbol{x}}^i} \right\}<\underline\epsilon_i(s_i^*)\right\}.
\end{align}
{By applying Theorem \ref{th:scenario} to every agent $i\in\{1,\ldots,N\}$, in the sense that it holds only for the the CBF constraints that involve agent $i$,} we have that for any $i\in\{1,\ldots,N\}$
\begin{align}\label{eq:scinequality2}
   &\mathbb{P}^{{ R}}\left\{\boldsymbol{x}\in{ \mathcal{H}}:\underline\epsilon_i(s_i^*)\le\mathbb{P}\left\{\vphantom{\bigcup}\boldsymbol{x}\in{ \mathcal{H}}:\boldsymbol{z}^*\notin{\mathcal{Z}_{\boldsymbol{x}}^i} \right\}\le\bar\epsilon_i(s_i^*)\right\}\nonumber\\
   &\ge 1-\beta_i\nonumber\\
   \Rightarrow&\sum_{i=1}^N\mathbb{P}^{{ R}}\left\{\bar\epsilon_i(s_i^*)<\mathbb{P}\left\{\boldsymbol{x}\in{ \mathcal{H}}:\vphantom{\bigcup}\boldsymbol{z}^*\notin{\mathcal{Z}_{\boldsymbol{x}}^i} \right\}\right.\nonumber\\
    &\bigcup\left.\mathbb{P}\left\{\boldsymbol{x}\in{ \mathcal{H}}:\vphantom{\bigcup}\boldsymbol{z}^*\notin{\mathcal{Z}_{\boldsymbol{x}}^i} \right\}<\underline\epsilon_i(s_i^*)\right\}<\sum_{i=1}^N\beta_i.
\end{align}
{Here $s_i^*$ is the number of $\boldsymbol{x}^{(r)}$'s for which there exists $e\in\mathcal{C}_i$, such that $\sum_{k\in\mathcal{V}_e}h_{ke}(u_k(\boldsymbol{x}^{(r)}))\ge \sum_{k\in\mathcal{V}_e}z_{ke}^*$. For a specific $r$, this means that agent $i$ recognizes that at least one CBF constraint is violated up to level $\sum_{i\in\mathcal{V}_e}z_{ie}^*$, over this scenario $\boldsymbol{x}^{(r)}$. After solving the scenario program \eqref{eq:dscenvarification} and communicating with the neighbouring agents in $\mathcal{G}_e$ in a distributed manner, every individual agent is able to compute $\overline{\epsilon}_i^*(s_i^*),\underline{\epsilon}_i^*(s_i^*)$ by \eqref{eq:scenariopolynomial}, \eqref{eq:scenariopolynomial2}.}
Since $\frac{\underline \epsilon(s^*)}{N}<\underline\epsilon(s^*)<\bar\epsilon(s^*)$,
substituting \eqref{eq:scinequality2} into \eqref{eq:scinequality1} with $i=1,\ldots,N$ we obtain 
\begin{equation}\label{eq:55}
  \mathbb{P}^{{ R}}\left\{\frac{\underline\epsilon(s^*)}{N}\le\mathbb{P}\left\{\boldsymbol{x}\in{ \mathcal{H}}:\boldsymbol{z}^*\notin{\mathcal{Z}_{\boldsymbol{x}}} \right\}\le\bar\epsilon(s^*)
  \right\}\ge 1-\beta.
\end{equation}
% We then prove that $\boldsymbol{z}^*$ is unique. For the case where all the CBF constraints are satisfied, i.e. $\sum_{k\in\mathcal{V}_e} h_{ke}(u_k(\boldsymbol{x}^{(r)}))\le 0,\forall e=1,\ldots,E$, $r=1,\ldots,{ R}$, we have that $\boldsymbol{z}^*=0$ and $\boldsymbol{\zeta}^*= 0$. For the case where there exists violated CBF constraint, i.e. $\sum_{i\in\mathcal{V}_e}h_{ie}(u_i(\boldsymbol{x}^{(r)}))>0$, we have that $z_{ie}^*=0$ since $\boldsymbol{z}\le 0$, and $\zeta_{ie}^*>0$ for $i\in\mathcal{V}_e$. In summary, we always have $\boldsymbol{z}^*=0$ for any scenarios, thus \eqref{eq:55} is equivalent to \eqref{eq:violation}. In addition, we directly obtain that $\zeta_{ie}^*> 0$ shows that $\sum_{i\in\mathcal{V}_e}h_{ie}(u_i(\boldsymbol{x}^{(r)}))> z_{ie}^*=0$. Thus, for every agent, $s_i^*$ is the number of non-zero $\boldsymbol{\zeta}_{i}^*$.

\end{pf}

\bibliographystyle{ieeetr}
\bibliography{main.bib}

\newpage
\section*{Biographies}

\noindent
\begin{wrapfigure}{l}{1in}
    \centering
    \includegraphics[width=1in,height=1.2in,clip,keepaspectratio]{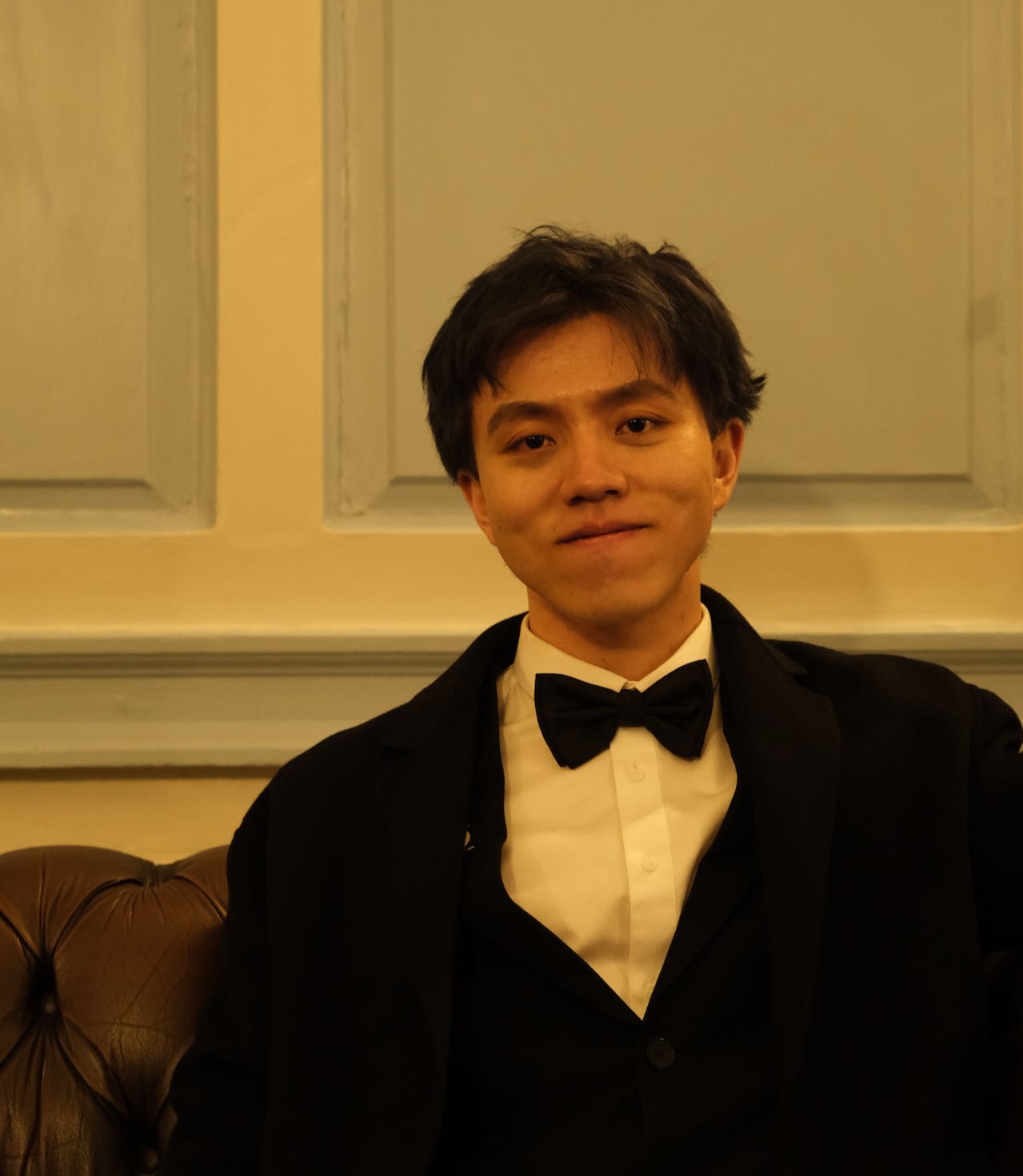}
\end{wrapfigure}
\noindent
\textbf{Han Wang} received the B.S. degree in Information Security from Shanghai Jiao Tong University, China, in 2020, and the Ph.D. degree in control from the the University of Oxford, UK, in 2024. He is currently a Postdoctoral Researcher at ETH Zürich. His research interests include safe and stable control design, learning to control and convex optimisation, with applications to industrial automation.

\vspace{4mm}

\noindent
\begin{wrapfigure}{l}{1in}
    \centering
    \includegraphics[width=1in,height=1.25in,clip,keepaspectratio]{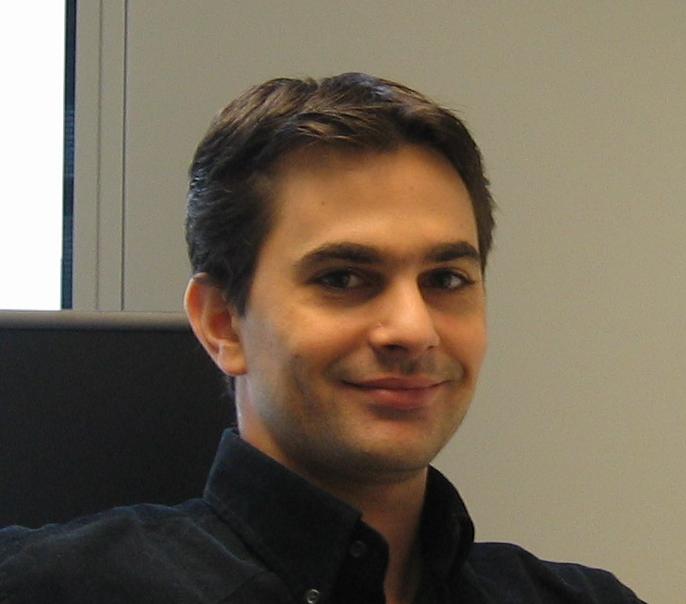}
\end{wrapfigure}
\noindent
\textbf{Antonis Papachristodoulou} (Fellow, IEEE) received the M.A./M.Eng. degree in Electrical and Information Sciences from the University of Cambridge, United Kingdom, and the Ph.D. degree in Control and Dynamical Systems (with a minor in Aeronautics) from the California Institute of Technology, Pasadena, CA, USA. He is currently a Professor of Engineering Science at the University of Oxford, United Kingdom, and a Tutorial Fellow at Worcester College, Oxford. He was previously an EPSRC Fellow. His research interests include large-scale nonlinear systems analysis, sum-of-squares programming, synthetic and systems biology, networked systems, and flow control.

\vspace{4mm}

\noindent
\begin{wrapfigure}{l}{1in}
    \centering
    \includegraphics[width=1in,height=1.25in,clip,keepaspectratio]{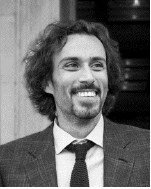}
\end{wrapfigure}
\noindent
\textbf{Kostas Margellos} received the Diploma degree in Electrical Engineering from the University of Patras, Greece, in 2008, and the Ph.D. degree in Control Engineering from ETH Zürich, Switzerland, in 2012. From 2013 to 2015, he was a Postdoctoral Researcher with ETH Zürich, UC Berkeley, USA, and Politecnico di Milano, Italy. In 2016, he joined the Control Group, Department of Engineering Science, University of Oxford, United Kingdom, where he is currently an Associate Professor. He is also a Fellow of Reuben College, Oxford, and a Lecturer at Worcester College, Oxford. His research interests include optimization and control of complex uncertain systems, with applications to energy and transportation networks.
\end{document}